\documentstyle[preprint,aps,epsf]{revtex}

\def\by#1#2{{\displaystyle {#1}\over \displaystyle {#2}}}
\def\d{{\rm d}}
\preprint {\begin{tabular}{l} IMSc/2001/01/01
\\ {MRI-P-010101} \\ \end{tabular}}
\begin{document}
%\draft
\title{Neutrinos from Stellar Collapse: \\
Comparison of signatures in water and heavy water detectors}

\author{G. Dutta,}
\address{Harish-Chandra Research Institute, Allahabad 211 019, India,}
\author{D. Indumathi, M. V. N. Murthy and G. Rajasekaran}
\address
{The Institute of Mathematical Sciences, Chennai 600 113, India.
}
\date{\today}
\maketitle
\begin{abstract}

Signatures of neutrino and antineutrino signals from stellar collapse
in heavy water detectors are contrasted with those in water detectors.
The effects of mixing, especially due to the highly dense matter in the
supernova core, are studied. The mixing parameters used are those sets
allowed by current understanding of available neutrino data: from
solar, atmospheric and laboratory neutrino experiments. Signals at a
heavy water detector, especially the dominant charged current
reactions on deuteron, are very sensitive to some of these sets of
allowed mixing parameters. Theoretical uncertainties on supernova
neutrino spectra notwithstanding, a combination of supernova
measurements with water and heavy water detectors may be able to
distinguish many of these mixing possibilities and thus help in ruling
out many of them.

\end{abstract}


\narrowtext

\section{Introduction}

Neutrinos from stellar collapse have so far been detected only from the
supernova SN1987a in the Large Magellanic Cloud\cite{sn1987a}. The
inital observations of the neutrino and antineutrino events by the
Kamiokande\cite{kii} and IMB\cite{imb} collaborations were the subject of
detailed analysis by several authors
\cite{dass,arafune,bahcall,sato,arnett,kolb,cowsik} immediately following
the event. The analyses confirmed the qualitative features of core
collapse and subsequent neutrino emission. 

The effect of non-zero neutrino masses and mixing on supernova signals
was first analysed in general by Kuo and Pantaleone\cite{KuoP}.
Recently, several authors have looked at the possible signatures of
neutrinos and antineutrinos from supernova collapse in realistic
scenarios. Dighe and Smirnov\cite{dighe} have looked at the problem of
reconstruction of the neutrino mass spectrum in a three-flavour scenario.
These authors as also Chiu and Kuo\cite{chiu} have compared the
signatures in the standard mass hierarchy and inverted mass hierarchy.
While these papers incorporate the constraints from solar and
atmospheric neutrino observations, the important question of the mass
limits that may be obtained from the observation of time delay has been
analysed by Beacom and Vogel\cite{beacom1,beacom2} and by Choubey and
Kar\cite{choubey} (see also the review by Vogel\cite{vogel}).  For a
recent review which also discusses aspects of locating a supernova by
its neutrinos in advance of optical observation, see
Ref.~\cite{beacom}.

In this paper we apply the analysis of neutrinos from stellar collapse
presented in Refs.~\cite{dutta1,dutta2} to heavy water detectors.  We
had previously discussed in detail the signatures, in a water Cerenkov
detector, of neutrinos and antineutrinos from stellar collapse in both
3- and 4-flavour mixing scenarios. The 4-flavour analysis was motivated
by {\sc lsnd} data \cite{lsnd}. The analysis was confined to Type II
supernovae (which occur when the initial mass of the star is larger than
8 solar masses). The choices of mixing parameters used were
consistent with available data on solar, atmospheric and laboratory
neutrino experiments. It turns out that different choices of (allowed)
mixing parameters lead to drastically different supernova signals at
water detectors. While certain features are common to both the 3- and
4-flavour analyses, there are important differences that may (depending
on the mixing parameters) be able to distinguish the number of
flavours. We will summarise the salient features of this analysis
below.

The dominant contribution is from the charged current (CC) scattering
of $\overline{\nu}_e$ on protons in water. Mixing can increase the
number of high energy events in this channel.  The most dramatic effect
of 3-flavour mixing is to produce a sharp increase in the CC events
involving oxygen targets \cite{dutta1,haxton} for most choices of mixing
parameters. These will show up as a marked increase in the number of
events in the backward direction with respect to the forward peaked
events involving electrons as targets (more than 90\% of which lie in a
$10^\circ$ forward cone with respect to the supernova direction for
neutrinos with energies $E_\nu~ {\stackrel{{}_{\displaystyle
>}}{{}_{\displaystyle \sim}}}~8$ MeV), both showing up over the mostly
isotropic CC events on protons.  In the absence of such mixing, there
will be only a few events due to CC scattering on oxygen targets. These
events involving oxygen targets will be seen in a heavy water detector
as well; realistically however, there will be fewer such events due to
the considerably smaller size of the heavy water detector at the
Sudbury Neutrino Observatory ({\sc sno}) as compared to the water
detector at SuperKamiokande (SuperK).

When 4-flavour mixing is considered, the analysis becomes obviously
more complex; now, the increase due to oxygen events will be visible
only for some of the allowed values of the parameters. Furthermore,
there is no set of allowed parameters in both the 3- and 4-flavour
cases for which the signals in a water detector will be able to
distinguish between adiabatic and non-adiabatic neutrino propagation in
the core of the supernova. This is an important issue since it can place
a lower bound on the (13) mixing angle, which is currently bounded by
{\sc chooz} at the upper end in the case of 3-flavour mixing. (The only
non-zero value for this angle so far comes from the {\sc lsnd} data
which can only be analysed in a 4-flavour frame-work). A zero value for
this mixing angle will decouple the two sets of oscillations, $\nu_e \to
\nu_\mu$ and $\nu_\mu \to \nu_\tau$, and allow for a simple 2-flavour
analysis separately of the solar and atmospheric neutrino data. An
exactly zero value of this angle will also render irrelevant CP
violating phases in the problem.

There are also CC and neutral current (NC) events due to elastic
scattering with electrons in water/heavy water. These interactions are
the same in both detectors and have been discussed in detail in the
earlier analysis on water detectors \cite{dutta2}.

In this paper we focus on the possible signatures of a supernova
collapse in a heavy water detector. The question is of practical
interest since {\sc sno} is operating for more than a year now. It
turns out that a combination of measurements in water and heavy water
detectors has much better discriminatory power than either of them
individually. Hence such observations of supernova neutrinos may be a
good signal to rule out some of the currently allowed parameter space
in the neutrino mixing angles. (These signals are however not very
sensitive to neutrino mass squared differences.)

The most interesting signals to study at a heavy water detector are the
events from CC interactions of both $\nu_e$ and $\overline{\nu}_e$ on
deuterons. This is the most dominant signal in contrast to the dominant
CC interaction of $\overline{\nu}_e$ alone on free protons in a water
detector (all are typically about two orders of magnitude larger than
those from oxygen or electron interactions). We will focus mostly on
these events in this paper, besides making a few remarks on the NC
events on deuteron that can also easily be measured at a heavy water
detector.  These are therefore the ``new'' signals in a heavy water
detector that would not be observable in (or will be different from) a
water detector.

As before the analysis is done assuming the standard mass hierarchy
necessitated by the solar and atmospheric neutrino observations
\cite{snatm}. We also impose all the known constraints on the mixing
parameters and mass-squared differences including those constraints
from laboratory experiments\cite{lsnd,chooz}. The purpose of the
calculation is to see whether different mixing scenarios give
significantly different signals in the detector.

In Sect. II we briefly outline the frame-work including the mixing
matrix as also matter effects on mixing. We then use this to obtain
expressions for the $\nu_e$, $\overline {\nu}_e$ and $\nu_{\mu,\tau}$,
$\overline {\nu}_{\mu,\tau}$ fluxes reaching the detector in both 3-
and 4-flavour mixing schemes. While discussions may be found in
Ref.~\cite{dutta1,dutta2} we reproduce the relevant details to keep
this paper self-contained. In Sect. III we list the different
interaction processes relevant to both water and heavy water detectors,
in particular, those with electrons (positrons) in the final state, as
well as NC events on deuteron.  The details of the supernova model used
in the calculation as well as the allowed values of neutrino mixing
parameters (from already existing data) using which the effects of
mixing are computed are also listed here. Results showing the effects
of mixing both for the spectrum and the integrated number of events are
presented in Sect. IV. Sensitivity to the supernova model parameters is
discussed. The NC events on deuterons in a heavy water
detector are also briefly discussed here. All these results are for a
$(2+2)$ mass hierarchy scheme in the case of 4 flavour mixing. In Sect.
V, we briefly discuss the other possible scenario, that of the $(3+1)$
scheme in 4-flavours. We present a summary and discussions in Sect. VI.

\section{Neutrino mixing and matter effects}

We briefly review mixing among three and four flavours of neutrinos (or
antineutrinos) and compute the neutrino (antineutrino) survival and
conversion probabilities. These are given in more detail in
Refs.\cite{dutta1,dutta2}. The supernova neutrinos are produced mostly
in the core, where the matter density is very high. Hence matter
effects on the propagation are important.

\begin{figure}[htp]
\vskip 8truecm
\includegraphics{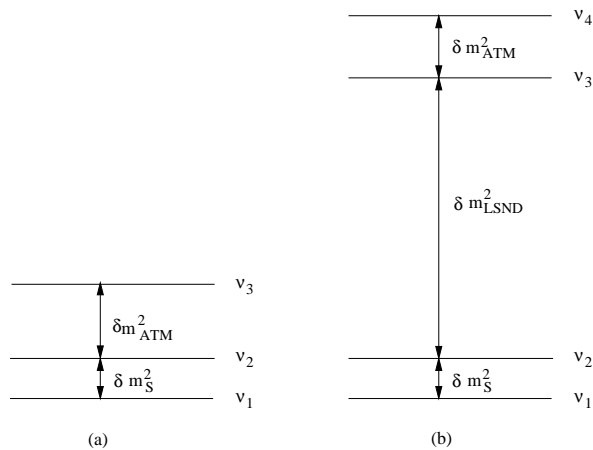}
\caption{The vacuum mass square differences in the 3 and 4 flavour
schemes. In the 4-flavour scheme, $\nu_e$ and $\nu_s$ are
predominantly mixed states of $\nu_1$ and $\nu_2$ while $\nu_\mu$ and
$\nu_\tau$ are that of $\nu_3$ and $\nu_4$ (2+2 scheme).  The mixing
between the lower and upper doublets has been chosen to be very small.
Here {\sc s, atm and lsnd} stand for the solar, atmospheric and {\sc
lsnd} mass squared differences respectively.}
\label{fig:level} % fig1
\end{figure}

The hierarchy of mass eigenstates is shown in Fig.~\ref{fig:level}. The
3-flavour case is shown in Fig.~\ref{fig:level}a; there are essentially
two scales corresponding to the solution of solar and atmospheric
neutrino problems assuming neutrino oscillations. In the case of four
flavours (necessitated by the non-zero result of {\sc lsnd}
\cite{lsnd}), the analysis is more complicated since there is an
additional sterile neutrino. One possible mass hierarchy here is a
$(2+2)$ scenario \cite{Pakvasa} as shown in Fig.~\ref{fig:level}b. The
{\sc lsnd} result implies the existence of a mass scale in the range of
0.1 eV$^2$ to 1 eV$^2$. We choose two doublets separated by this mass
scale. The intra-doublet separation in the lower doublet corresponds to
the solar neutrino scale $~{\stackrel{{}_{\displaystyle
<}}{{}_{\displaystyle \sim}}}~10^{-5}$ eV${}^2$ and that in the upper
one to the atmospheric neutrino scale $\sim 10^{-3}$ eV${}^2$.  Yet
another possibility is the so-called $(3+1)$ scheme \cite{MRS,PS},
where the 3-flavour scheme is extended by adding a heavy sterile
neutrino as the heaviest mass eigenstate, with a mass separation from
the other three states as required by {\sc lsnd}. We discuss this scheme
separately later. In what follows, therefore, 4-flavour mixing always
refers to the $(2+2)$ scheme.

The mixing matrix, $U$, which relates the flavour and mass eigenstates
in the above scenarios has three angles in the 3-flavour case and six
angles in the 4-flavour one. We shall ignore the CP violating phases.
Then, the mixing matrix can be parametrised in the case of 3-flavours as,
\begin{equation}
\left[ \nu_e \quad \nu_\mu \quad \nu_\tau \right ]^T = U \times 
\left[ \nu_1 \quad \nu_2 \quad \nu_3 \right ]^T~,
\end{equation}
where $T$ stands for transpose. Here $U$ is parametrised by considering
rotations of mass eigenstates taken two at a time:
\begin{eqnarray}
U & = & U_{23} \times U_{13} \times U_{12}~, \nonumber
\\
  & = & U_\psi \times U_\phi \times U_\omega~,
\end{eqnarray}
where $U_{ij}$ corresponds to the rotation of mass eigenstates $\vert i
\rangle$ and $\vert j \rangle$.  As is the convention, we denote the
mixing angle relevant to the solar neutrino problem by $\omega$ and
that relevant to the atmospheric neutrino problem by $\psi$ in the case
of 3 flavours. The (13) angle $\phi$ is then small, due to the {\sc chooz}
result \cite{chooz} which translates to the constraint on the (13)
mixing angle, $\sin \phi \equiv \epsilon \le \epsilon_0$, where
$\epsilon_0 = 0.16$. In the case of 4-flavours, we choose to work in
the $(2+2)$ scheme:

\begin{equation}
\left[ \nu_e \quad \nu_s \quad \nu_\mu \quad \nu_\tau \right ]^T = U \times 
\left[ \nu_1 \quad \nu_2 \quad \nu_3 \quad \nu_4 \right ]^T~,
\end{equation}
where the mixing matrix is similarly defined to be,
\begin{eqnarray}
U & = & U_{34} \times U_{24} \times U_{23} \times
U_{14} \times U_{13} \times U_{12}~, \nonumber \\
  & = & U_\psi \times U_\epsilon \times U_\epsilon \times 
  U_\epsilon \times U_\epsilon \times U_\omega~.
\end{eqnarray}
Here the (13) and (14) angles are constrained by {\sc chooz} to be
small: $\theta_{13},\theta_{14} \sim \epsilon \le \epsilon_0$
\cite{dutta2}. The
atmospheric neutrino problem constrains $\theta_{34}$ (the equivalent
of the angle $\psi$ in the 3-flavour case) to be maximal, when both
$\theta_{23},\theta_{24}$ become small. We shall assume they are also
limited by the same small parameter, $\epsilon_0$.

The other relevant definition is the chosen mass hierarchy in the
problem. We define the mass squared differences as $\delta_{ij} =
\mu_i^2 - \mu_j^2$, where $i,j$ run over the number of flavours.
Without loss of generality, we can take $\delta_{21}$, $\delta_{31}$
(and $\delta_{41}$) to be greater than zero; this defines the standard
hierarchy of masses consistent with the range of the mixing angles, as
specified above.

\subsection{Matter effects for neutrinos}

\begin{figure}[htp]
\vskip 8truecm
{\includegraphics{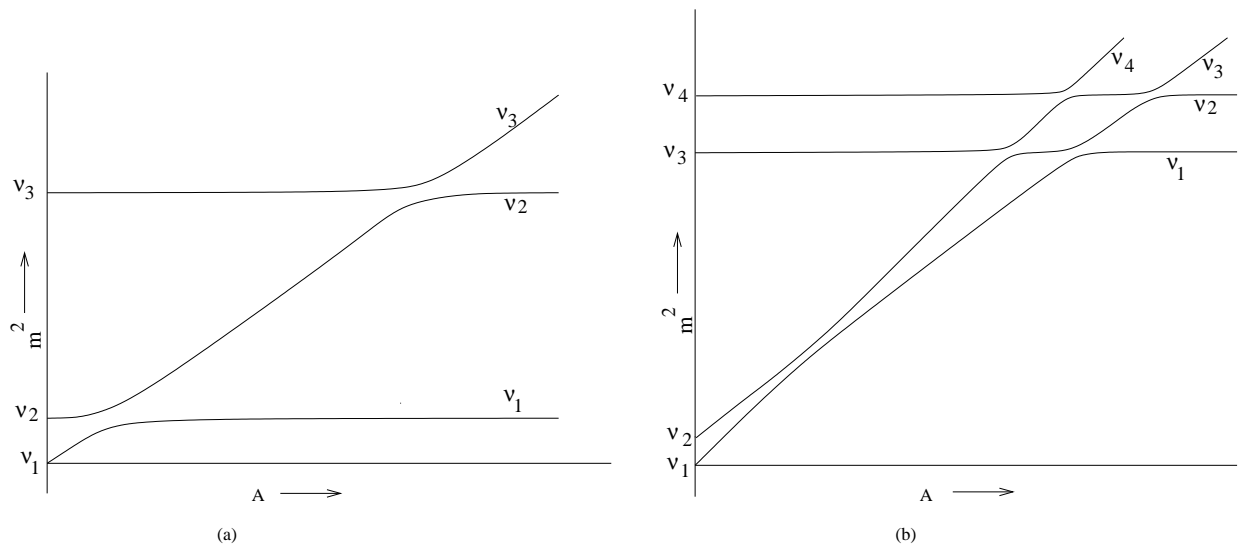}}
\caption{Schematic drawing showing mass squares as functions of matter
density in the 3 and 4 flavour schemes. Resonances occur at two
different regions of matter density, the lower one at $\approx \delta
m^2_S$. In the 4-flavour case the upper resonances consist of four
close-lying resonances determined by $\delta m^2_{\sc lsnd}$.}
\label{fig:nu34} % fig 2
\end{figure}

The most important consequence of the highly dense core matter is to
cause Mikheyev, Smirnov and Wolfenstein (MSW) resonances \cite{MSW} in
the neutrino sector. (The chosen mass hierarchy prevents such
resonances from occurring in the antineutrino sector). In both the 3-
and 4-flavour cases, this causes the electron neutrino to occur
essentially as a pure mass state \cite{dutta2}, in fact, as the highest
mass eigenstate, as can be seen from the schematic illustration in
Fig.~\ref{fig:nu34}. Nonadiabatic transitions near the resonances will
alter this result, especially in the case of 4-flavours, where there
are several level crossings because of the presence of the sterile
neutrino. Hence we will discuss the purely adiabatic and the
non-adiabatic cases separately.

\subsubsection{The adiabatic case}
We begin with the 3-flavour case. The average transition probability
from flavour $\beta$ to flavour $\alpha$ is denoted by $P_{\alpha
\beta}$ where $\alpha,\beta$ run over all flavours. It turns out that,
independent of other mixing angles and the mass squared differences,
the survival probability for the adiabatic case is,
\begin{equation}
P_{ee}=\epsilon^2~,
\end{equation}
and hence is small. This is sufficient to determine all the relevant
fluxes in the 3-flavour case. In the 4-flavour adiabatic case, we again
have,
\begin{equation}
P_{ee}=\epsilon^2~,
\end{equation}
so that the electron neutrino survival probability is flavour
independent in the adiabatic case. In addition, we also need to know
some other probabilities in order to compute the fluxes at the
detector. We have
\begin{equation}
P_{es} = P_{se} = P_{ss} = P_{ee}=\epsilon^2~.
\end{equation}
%In the $(3+1)$ 4-flavour scheme, however, these probabilities are
%different and depend on the (34) mixing angle, which is unconstrained.
%We therefore focus hereafter on the $(2+2)$ 4-flavour scheme with only
%a passing reference to this scheme when we discuss the NC events.

\subsubsection{The non-adiabatic case}
Because of the parametrisation it is easy to see that non-adiabatic
effects in the form of Landau Zener (LZ) jumps are introduced as a
result of the values chosen for $\epsilon$ and $\omega$. The value of
$\epsilon$ determines whether non-adiabatic jumps are induced at the
upper resonance(s) while the value of $\omega$ determines whether the
non-adiabatic jump occurs at the lower resonance. This statement holds
both for three and four flavours since in both cases the
non-adiabaticity in the upper resonance(s) is controlled by $\epsilon$,
apart from mass squared differences.

For a large range of $\epsilon$, allowed by the {\sc chooz} constraint,
the evolution of the electron neutrino is adiabatic. As a result the
lower resonance does not come into the picture at all except when
$\epsilon$ is very close to zero, where the LZ jump probability at the
upper resonance(s), $P_H$, abruptly changes to one \cite{KuoP,LZ}. This
occurs in a very narrow window: for instance, when $\epsilon$ changes
from 0.02 to 0.01 $P_H$ changes from 0.01 to 0.34 for a mass scale in
the range of $10^{-3}$ eV${}^2$. The subsequent discussion is therefore
relevant only when $\epsilon$ is vanishingly small, $\epsilon \ll
10^{-2}$. (This is not ruled out by the known constraints except {\sc
lsnd}).

The LZ transition at the lower resonance is determined by the
probability, $P_L$, which is a function of the mixing angle $\omega$
and the mass squared difference $\delta_{21}$. It turns out that $P_L$
is zero unless $\omega$ is small,
$\sin\omega~{\stackrel{{}_{\displaystyle <}}{{}_{\displaystyle
\sim}}}~0.2$ for mass differences in the solar neutrino range $ \sim
10^{-5}$ eV${}^2$. This case (of small $\epsilon$, small $\omega$)
corresponds to the extreme non-adiabatic limit. When $\omega$ is large,
as in the case of the large-angle MSW or the vacuum solution to the
solar neutrino problem, non-adiabaticity occurs only at the upper
resonance and is therefore partial.

In our calculations we have used the form for $P_H$ and $P_L$
discussed\footnote{The Landau Zener transition probability is defined
in terms of the adiabaticity parameter, $\gamma$, as $P_{LZ} =
\exp[-(\pi/2) \gamma F]$, where $F \sim 1$. The final expression for
$\gamma$ given in Appendix B of Ref.~\cite{dutta1} should be multiplied
by the additional factor $\delta_{31}/60$. However, the numerical
calculation in that paper is correct.} in the appendix of
Ref.\cite{dutta1} (see also \cite{LZ}).  Because of the sharpness of
the transition at the upper resonance we will set $P_H = 1$ for small
enough $\epsilon$ and only consider the dependence of the survival
probability on the LZ transition at the lower resonance.

In the 3-flavour non-adiabatic case the only relevant probability is,
\begin{equation}
P_{ee}= (1 - \epsilon^2)[(1-P_L)\sin^2 \omega~ + P_L \cos^2\omega]~.
\end{equation}
In the non-adiabatic 4-flavour case we have,
\begin{eqnarray}
P_{ee}&=& (1-2 \epsilon^2)[(1-P_L)\sin^2 \omega~ + P_L \cos^2\omega]~,
\nonumber \\ 
P_{es}&=& (1-2 \epsilon^2)[(1-P_L)\cos^2 \omega~ + P_L \sin^2\omega]~.
\label{eq:nonad4}
\end{eqnarray}
Since $\epsilon$ is small, the flux at the detector is entirely
controlled by $\omega$ and $P_L$ which is also a function of $\omega$.
Note that the sum $P_{ee} + P_{es} = 1 - 2\epsilon^2$ is independent of
$P_L$ in 4-flavours. Since $P_{ee} + P_{es} = 1 - P_{e\mu} -
P_{e\tau}$, this also indicates that the probability of transition of
$\nu_{\mu,\tau}$ into $\nu_e$ is small in the non-adiabatic case, in
contrast to the adiabatic case where $P_{ee} + P_{es} = 2\epsilon^2$.

\subsection{Matter effects for antineutrinos}

Due to our choice of mass hierarchy, the propagation of antineutrinos is
always adiabatic (the matter dependent term has the opposite sign as 
compared to neutrinos). Here also the high density of matter in the core
causes the electron antineutrino to be produced as a pure mass
eigenstate, $\overline{\nu}_1$.

In the 3-flavour case, we have
\begin{equation}
P_{\overline{e}\overline{e}} = (1-\epsilon^2)\cos^2\omega~.\\ 
\label{eq:pbar3}
\end{equation}
In the 4-flavour case, we have,
\begin{eqnarray}
P_{\overline{e}\overline{e}} &=& (1-2 \epsilon^2)\cos^2\omega~,\nonumber\\ 
P_{\overline{e}\overline{s}} &=& (1-2 \epsilon^2)\sin^2\omega~.
\label{eq:pbar4}
\end{eqnarray}
Hence, when $\omega$ is small, there is very little loss of the
$\overline{\nu}_e$ flux into the sterile channel. Also, the sum
$P_{\overline{e}\overline{e}} + P_{\overline{e}\overline{s}} = (1- 2
\epsilon^2)$ is
similar to the non-adiabatic neutrino propagation so that very
little $\overline{\nu}_{\mu,\tau}$ is converted to $\overline{\nu}_e$.

For the NC events, we will also need the following:
\begin{eqnarray}
P_{\overline{s}\overline{e}} &=& (1-2\epsilon^2) \sin^2\omega + 2
\epsilon^2 \sin 2 \omega~, \nonumber \\
P_{\overline{s}\overline{s}} &=& (1-2\epsilon^2) \cos^2\omega - 2
\epsilon^2 \sin 2 \omega~.
\end{eqnarray}
These probabilities can then be used to determine the observed
antineutrino fluxes.

\subsection{The neutrino (antineutrino) fluxes at the detector}

Following Kuo and Pantaleone \cite{KuoP}, we denote the flux
distribution, $\d \phi_\alpha^0/\d E$, of a neutrino (or
antineutrino) of flavour $\alpha$ with energy $E$ produced in the core
of the supernova by $F_\alpha^0$. In particular we use the generic
label $F_x^0$ for flavours other than $\nu_e$ and $\overline\nu_e$
since
\begin{equation}
F_x^0 = F_{\overline{x}}^0
= F_{{\mu}}^0
= F_{\overline{\mu}}^0
= F_{{\tau}}^0
= F_{\overline{\tau}}^0~.
\end{equation}

Typically, models for supernovae predict that the $\overline{\nu}_e$
and $\nu_x$ have hotter spectra than $\nu_e$. This is
because the $\nu_x$ decouples first since it has only NC interactions
with matter and therefore leaves the cooling supernova with the hottest
thermal spectrum. The CC interactions of $\nu_e$ with matter are larger
than those of $\overline{\nu}_e$ and hence $\nu_e$ has the coldest
spectrum. The model on which this work is based \cite{Totani} predicts
that the average energies of the $\nu_e$, $\overline{\nu}_e$ and
$\nu_x$ spectra are around 11, 16, and 25 MeV. We make our observations,
keeping this in mind.

The $\nu_e$ flux on earth is given in terms of
the flux of neutrinos produced in the core of the supernova by,
\begin{eqnarray}
F_{e} & = & P_{ee} F_{e}^0 +P_{e\mu}F_{{\mu}}^0 + 
                P_{e\tau}F_{{\tau}}^0, \nonumber \\ 
      & = & P_{ee}F_{e}^0 +(1-P_{ee})F_x^0 \qquad \qquad
					   (\hbox{\rm 3-flavours)}~,
      \nonumber \\
      & = & P_{ee}F_{e}^0 +(1-P_{ee}-P_{es})F_x^0~\quad
					   (\hbox{\rm 4-flavours)}~,
\label{eq:fe}
\end{eqnarray}
where we have made use of the constraint $\sum_{\beta} P_{\alpha\beta}
=1$. This flux is further reduced by an overall geometric factor of
$1/(4\pi d^2)$ for the case of a supernova at a distance $d$ from the
earth.

Since the probabilities $P_{ee}$ and $P_{es}$ are known, the $\nu_e$
flux can be computed in terms of the original fluxes emitted by the
supernova.

The $\nu_e$ flux is independent of the (12) mixing angles $\omega$
in the adiabatic case. Also, it is not very different for 3- and
4-flavours. In both, the observed flux is almost entirely due to the
original $\nu_x$ flux since $P_{ee} = \epsilon^2$ is small, and is
therefore hotter.

From Eq.~(\ref{eq:nonad4}) we see that, in contrast, the contribution
from the hotter spectrum into electron neutrinos in the 4-flavour
non-adiabatic case is controlled entirely by $\epsilon$ and is small.
However, the observed flux can be depleted, depending on the value of
$P_L$. The signal in the 3-flavour case is drastically different
because the contribution of the hotter spectrum now depends on
$\epsilon$, $\omega$ and $P_L$. In general, the possibility of LZ
transitions makes the analysis more complicated. We will discuss this
case numerically later.

The result for $\overline{\nu}_e$ is the same, with $P_{ee}$
replaced by $P_{\overline{e}\overline{e}}$, etc. For example,
\begin{eqnarray}
F_{\overline{e}}
      & = & P_{\overline{e}\overline{e}}F_{\overline{e}}^0
      +(1-P_{\overline{e}\overline{e}})F_x^0 \qquad \qquad
					   (\hbox{\rm 3-flavours)}~,
      \nonumber \\
      & = & P_{\overline{e}\overline{e}}F_{\overline{e}}^0
      +(1-P_{\overline{e}\overline{e}}-P_{\overline{e}\overline{s}})F_x^0~\quad
					   (\hbox{\rm 4-flavours)}~.
\label{eq:febar}
\end{eqnarray}
There is hardly any mixing of the hotter spectrum into electron
antineutrinos in the 4-flavour case since $(1 -
P_{\overline{e}\overline{e}} - P_{\overline{e}\overline{s}}) = 2
\epsilon^2$ is small.  The extent of mixing in 3-flavours depends on
the value of $\omega$ through $P_{\overline{e}\overline{e}}$.  For small
$\omega$, there is very little change in the electron antineutrino flux
(see Eq.~(\ref{eq:pbar4})).

The results are summarised in Tables~\ref{tab:adnu} and
~\ref{tab:adnubar} for the adiabatic neutrino and antineutrino cases,
and in Table~\ref{tab:nonad} for non-adiabatic neutrino propagation.
Expressions for the other flavours, needed to compute the NC events,
are also given in these tables. For instance,
\begin{eqnarray}
2 F_x & = & F_{{\mu}} + F_{{\tau}}~, \nonumber \\ 
      & = & (P_{ee}+P_{es}+P_{se}+P_{ss})F_x^0 
                +(1-P_{ee}-P_{se})F_{e}^0~ \qquad \hbox{(4 flavours)}~.
\label{eq:fx} 
\end{eqnarray}
A similar expression holds for $F_{\overline{x}}$ with $P_{\alpha\beta}$
replaced by $P_{\overline{\alpha}\overline{\beta}}$.

While the neutral current (NC) combination, $(\nu_e + 2\nu_x +
\overline{\nu}_e + 2 \overline{\nu}_x)$, remains unaltered in
3-flavours, there may be actual loss of spectrum into the sterile
channel in the 4-flavour case. This should also be a good indicator of
the number of flavours involved in the mixing.

Water as well as heavy water detectors will be sensitive to all these
aspects of mixing. In the next section, we will discuss the inputs and
constraints, both from supernova models as well as current neutrino
experiments. These will then be used subsequently to predict
numerically supernova event rates.

\section{Inputs and Constraints}

We will concentrate mostly on the dominant CC and NC interactions of
$\nu_e$ and $\overline{\nu}_e$ on deuteron in heavy water. We will also
compare the CC interactions to those at a water detector, that is, to
$\overline{\nu}_e$ on protons. Interactions on electrons and oxygen
nuclei in water and heavy water detectors are the same, and have been
discussed in detail in Refs.~\cite{dutta1,dutta2}.

\subsection{Interaction processes and relevant formulae}
Both in water and heavy water detectors, the interactions we are mainly
interested in are of three types:
\begin{enumerate}
\item Dominant events: These
are due to the CC $\overline{\nu}_e \, p \to e^+ \, n$
interaction in water. In heavy water, there are contributions from both
$\overline{\nu}_e \, d \to e^+ \, nn$ and
${\nu}_e \, d \to e^- \, pp$ CC processes. At the neutrino energies
relevant to this discussion, the cross-sections for these processes, and
hence the number of events, are about two orders of magnitude larger
than any of the others.
\item Electron events: These are due to elastic scatterings of
$\nu_e$, $\overline{\nu}_e$, $\nu_x$, $\overline{\nu}_x$ on electron
targets in both water and heavy water.
\item Oxygen events: These arise from CC scattering of $\nu_e$
and $\overline{\nu}_e$ on oxygen nuclei in both detectors.
\end{enumerate}
In all these cases, the events are identified by detecting an electron
(or positron) in the final state. The electron events, although small in
number, are highly forward peaked \cite{Bahcallb}. The oxygen events
have a high energy threshold ($E_{\nu_e} > 15.4$ MeV; $E_{\overline{\nu}_e}
> 11.4$ MeV) and can only occur if there is substantial mixing of the
hotter $\nu_x$ spectrum with the $\nu_e$ or the $\overline{\nu}_e$ in
which case these events are highly backward peaked \cite{haxton}. Both
these may therefore be readily separated from the mostly isotropic
dominant events in water \cite{beacom3}. In the case of heavy water, the
angular distribution of both the dominant processes is well-known
\cite{beacom3,kubo,butler} and is approximately given by
$$
P(\theta) \propto 1 - \by{1}{3} \cos\theta~,
$$
at the energies of interest. Here $\theta$ is the electron (positron)
laboratory scattering angle. So there are typically twice as many events
in the backward direction as in the forward direction. This makes the
exclusive identification of oxygen events more difficult in a heavy
water detector.

A note on the nomenclature ``forward'' and ``backward''. We envisage the
angular dependence as being measured in typically six bins of $30^\circ$
each, the first bin corresponding to ``forward'' and the last one
corresponding to the ``backward'' events. This obviates the need for
very accurate angle measurements as well as takes into consideration
effects like electron-rescattering that may smear out the scattering
angle. This bin size is also typically what is available at
SuperKamiokande. 

Since the supernova signal is a
short and well-defined signal (lasting about 10 secs), it will be possible
to detect all these events over background (due to solar and other
radioactive processes). For instance, in the relevant energy region,
SuperKamiokande reports \cite{SuperKback} a background of 0.2
events/day/kton in a bin of 1 MeV of scattered electron energy from the
direction of the sun, and half that rate from all other scattering
angles. The corresponding number for SNO is not available, although the
radioactive background is expected to be small compared to the solar
signal \cite{SNOback}. Furthermore, since their angular distribution is
well-known, angular information on the final state electron ($e^-$ or $e^+$)
may allow us to separate out these three types of events.  The
detailed analysis of signals due to mixing in the forward and backward
event samples is given in Ref.\cite{dutta2} for a water Cerenkov
detector. These results also hold for a heavy water detector (apart
from a scaling factor of 0.9 due to the mass difference between water
and heavy water). We will not discuss this further and concentrate on
the new results for the CC events on deuterons. In addition, there are
NC processes on the deuteron in heavy water that may be detected by the
subsequent neutron capture and the associated gamma rays.

The specific processes we will consider, for water and heavy water
detectors, therefore are
\begin{equation}
\overline{\nu}_e + p \to e^+ + n~,
\label{cr1}
\end{equation}
and
\begin{eqnarray}
\overline{\nu}_e + d & \to & e^+ + n +n~, \nonumber \\
\nu_e + d & \to & e^- + p + p~.
\label{cr2}
\end{eqnarray}
Common to both detectors are the processes, 
\begin{eqnarray}
\nu_\alpha + e & \to & \nu_\alpha + e \qquad
(\alpha = e, x, \overline{e}, \overline{x})~,
\label{cre}
\\
\nu_e + {}_8^{16}O & \to & e^- + X~, \nonumber \\
\overline{\nu}_e + {}_8^{16}O & \to & e^+ + X~,
\label{crO}
\end{eqnarray}
where the elastic scattering on electrons involves both CC and NC
interactions.

We will also discuss the interesting possibility of observing NC events
on deuteron in a heavy water detector:
\begin{eqnarray}
\overline{\nu} + d & \to & \overline{\nu} + n + p~, \nonumber \\
\nu + d & \to & \nu + n + p~.
\label{cr3}
\end{eqnarray}
The cross-sections for Eqs.~(\ref{cr1})--(\ref{cr3})
are well-known \cite{Bahcallb,crossd,haxton}. The $\overline{\nu}_e p$
cross-section is large in water Cerenkov detectors, being proportional
to the square of the antineutrino energy. In terms of total number of
events, therefore, water Cerenkov detectors are mostly dominated by
$\overline{\nu}_e\, p$ events. The deuteron CC
cross sections are comparable though the $\nu_e$ CC reaction on
deuteron has a slightly lower threshold and a somewhat larger
cross-section than the $\overline{\nu}_e$ CC. Also, the
$\overline{\nu}_e$ CC cross-section in heavy water is about 4 times
smaller than the corresponding one in water due to Pauli suppression.

When the recoil electron (positron) is detected, the time integrated
event rate due to neutrinos (or antineutrinos) of flavour $\alpha$ and
energy $E$ on target $T$, as a function of the recoil electron (or
positron) energy, $E_e$, is as usual given by,
\begin{equation}
\by{\d N_\alpha^{\rm T}(E_e)}{\d E_e} = \by{n_T}{4\pi d^2} \sum_b \Delta
                   t_b \int \d E F_\alpha (b) \by{\d
		   \sigma^{\rm T}}{\d E_e}~.
\label{eq:rate}
\end{equation}
The index $b$ refers to the time interval within which the (original)
thermal neutrino spectrum can be assumed to be at a constant
temperature $T_b(\alpha)$. Here $n_T$ refers to the number of scattering
targets (of $d$, $p$, $e$, or $O$) that are available in the detector.
Also, for processes involving $d$ and $p$, the hadron recoil is so small
that we assume
$E_e = E - \delta_{IF}$, where $\delta_{IF}$ is the mass difference
between the initial and final hadrons (including the binding energy).
While this results in a small threshold of a few MeV in these cases, the
threshold for oxygen processes is greater than 10 MeV. 
The total number of events from a given flavour of neutrino in a given
bin, $k$, of electron energy (which we choose to be of width 1 MeV)
then is
\begin{equation} N_\alpha^{T}(k) = \int_k^{k+1} \d E_e \by{\d
N_\alpha^{\rm T}}{\d E_e}~. 
\end{equation}

In the case of NC events in heavy water, the total
number of events is calculated according to
\begin{equation}
N_\alpha^{\rm NC} = \by{n_t}{4\pi d^2} \sum_b \Delta t_b
\int \d E F_\alpha (b) \sigma^{\rm NC}(E)~.
\end{equation}
Here again the cross-section is well-known \cite{crossd}.

\subsection{The supernova flux inputs}

As in \cite{dutta1,dutta2}, we compute the time integrated event rate
at prototype 1 kton water and heavy water detectors from neutrinos
emitted by a supernova exploding 10 kpc away. Results for any other
supernova explosion may be obtained by scaling the event rate by the
appropriate distance to the supernova and the size of the detector, as
shown in Eq.~(\ref{eq:rate}). We assume the efficiency and resolution
of the detectors to be perfect; this will only slightly enhance the
event rates near the detector threshold \cite{dutta2}.

We use the luminosity and average energy distributions (as functions of
time) for neutrinos of flavour $\alpha$ and energy $E$ as given in
Totani et al. \cite{Totani}, based on the numerical modelling of Mayle,
Wilson and Schramm \cite{BL}. In a short time interval, $\Delta t_b$,
the temperature can be set to a constant, $T_{b}(\alpha)$. Then the
neutrino number flux can be described, in this time interval, by a
thermal Fermi Dirac distribution,
\begin{equation}
F_\alpha^0(b) = N_0 \by{{\cal L}_b (\alpha)}{T_{b}^4(\alpha)}
		       \by{E^2}{(\exp(E/T_{b}(\alpha)) +1)} ~,
\end{equation}
at a time $t$ after the core bounce. Here $b$ refers to the time-bin,
$t = t_0 + b \Delta t$. We set the time of bounce, $t_0 = 0$. The
overall normalisation, $N_0$, is fixed by requiring that the total
energy emitted per unit time equals the luminosity, ${\cal L}_b
(\alpha)$, in that time interval. The precise values for ${\cal L}_b
(\alpha)$ and $T_{b}(\alpha)$ are taken from Ref.~\cite{Totani}. The
total emitted energy in all flavours of neutrinos is about $2.7 \times
10^{53}$ ergs. The general features of the model are as follows: The
temperature is roughly constant over the entire period of emission
(lasting roughly 10 seconds). Typical values are $T_{b}(\alpha) = 3.15
\langle E \rangle_\alpha$, with the average energy of each flavour,
$\langle E \rangle_\alpha = 11, 16, 25$ MeV for $\alpha = \nu_e,
\overline{\nu}_e$ and $\nu_x$ respectively. Also, the total emitted
energy is more or less equally distributed in all flavours. (The
luminosity of $\nu_e$ is higher than that of other flavours at early
times, while that of $\nu_x$ is higher after 1 sec). The number of
neutrinos emitted in each flavour, however, is not the same since their
average energies are different. While we use the values for the time
dependent temperature and luminosity as given by Ref.~\cite{Totani} in
our analysis, we also examine the effects due to possible variations of
these parameters and hence the sensitivity of our results to the
details of the supernova model used.

With large matter effects present in both the neutrino and antineutrino
sector, the validity of the average energies of $\nu_e$,
$\overline{\nu}_e$ and $\nu_x$ as 11, 16 and 25 MeV, respectively,
which were calculated without mixing, may be questioned. This is
especially so because re-scattering effects involve the flavour states
which may then equilibrate at different temperatures. We however note
that the highly dense matter projects the $\nu_e$ and
$\overline{\nu}_e$ states as almost pure mass eigenstates. Hence
thermalisation is not affected by effects of mixing. In fact, the
effects of mixing are significant only when the resonant densities are
reached, when the MSW effect can mix different flavour states. For the
parameter values as allowed from current neutrino data, this occurs
only outside the neutrinosphere ($R \sim 10^4$ km), and not at
the core ($R \sim 50$ km) where most of the neutrinos are produced.
This mixing therefore occurs between spectra which are already
thermalised with the above-mentioned temperatures. In the case of
$\nu_\mu$ and $\nu_\tau$ this argument does not go through. In the
highly dense core, these are mixtures of more than one mass
eigenstate. However, both mix only into each other and scatter through
exactly the same processes. Hence their temperatures also remain the
same as in the no-mixing case.

\subsection{The mixing parameters}

We impose the following known constraints on the mixing matrix in
vacuum both for three and four flavour scenarios. Consistent with the
{\sc chooz} constraint, namely $\sin\phi \sim \epsilon \le \epsilon_0 =
0.16$ which we have imposed at the level of the parametrisation itself,
we choose $\epsilon = 0.08$ for the 3-flavour adiabatic case and
$\epsilon = 10^{-4} \sim 0$ for the non-adiabatic case. In the case of
the 4-flavour scheme, we set $\theta_{13}, \theta_{14},
\theta_{23}, \theta_{24} = \epsilon$.

The constraint from the atmospheric neutrino analysis implies that the
relevant angle $\psi (= \theta_{34} ) \approx
\pi/4$, is near maximal and the relevant mass squared difference is of
the order of $10^{-3}$ eV${}^2$. Neither of these constraints directly
enter our calculations except to determine whether the upper resonance
is adiabatic or not depending on the value of $\epsilon$ as constrained
by the {\sc chooz} findings. We consider both possibilities here.

The value of the (12) mixing angle, $\omega$, is not yet known.
Combined data on solar neutrinos give three possible values
\cite{Bahcallr} :
\begin{enumerate}
\item $\sin^22\omega = 6.0\times 10^{-3}, \delta_{21} = 5.4 \times 
10^{-6}$ eV${}^2 $ (SMA). The small angle MSW solution.
\item $\sin^22\omega = 0.76, \delta_{21} = 1.8 \times 
10^{-5}$ eV${}^2 $ (LMA). The large angle MSW solution.
\item $\sin^22\omega = 0.96, \delta_{21} = 7.9 \times 
10^{-8}$ eV${}^2 $ (LMA-V). The large angle vacuum solution. 
\end{enumerate}
This has been slightly modified \cite{GG,GMPV} in view of new data from
SuperK; however, it remains true, in general, that $\omega$ may be small or
large, with $\delta_{21} \le 10^{-5}$ eV${}^2$. 

We will now discuss the results numerically for all these choices.

\section{Numerical Results}

The numerical calculations are done by following the evolution of the mass
eigenstates through all the resonances including the appropriate jump
probabilities when the transition is non-adiabatic (when $\epsilon$ is
small). Furthermore, the LZ jump at the lower resonances is significant
only for small values of $\omega$. 

\subsection{The electron (positron) spectrum}

\begin{figure}[htp]
\vskip 9truecm
{\includegraphics{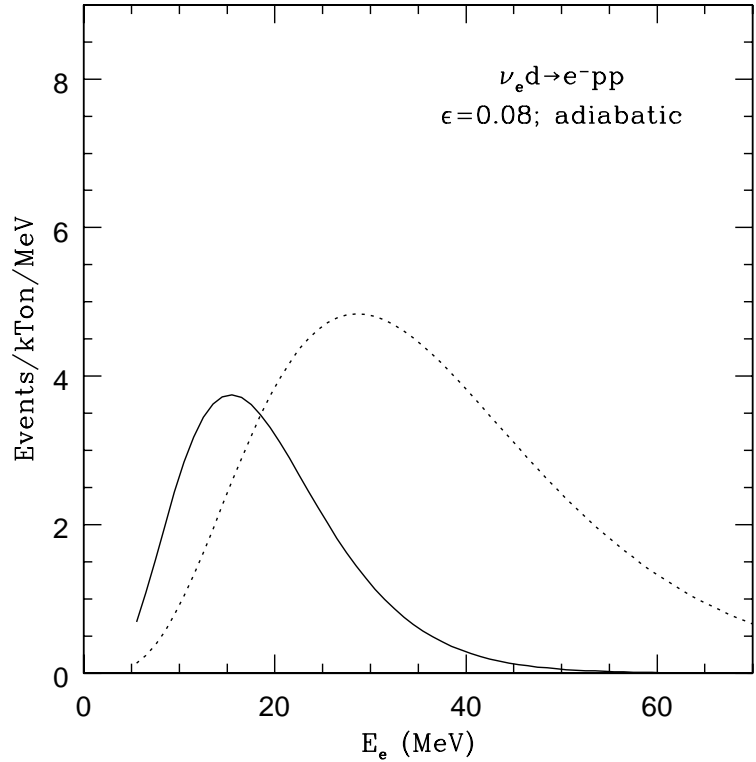}}
\caption{$\nu_e\, d$ event rates as a function of the scattered electron
energy $E_e$, when the upper resonance is completely
adiabatic. The solid line represents the no-mixing case. The dotted
line is due to the effects of either 3- or 4-flavour mixing,
which cannot be distinguished here.}
\label{fig:nuead} % fig 3

\end{figure}

\begin{figure}[htp]
\vskip 9.3truecm
{\includegraphics{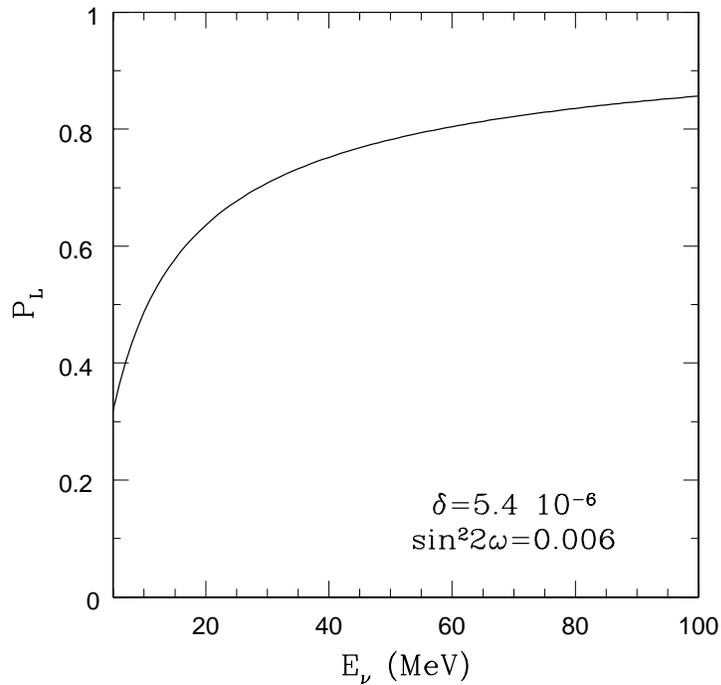}}
\caption{The Landau Zener jump probability $P_L$ at the lower resonance
as a function of the neutrino energy for the small angle MSW
solution (SMA) values of the mass squared difference and mixing angle
$\omega$ for the 3-flavour case.}
\label{fig:PL} % fig 4
\end{figure}

\begin{figure}[htp]
\vskip 8truecm
{\includegraphics{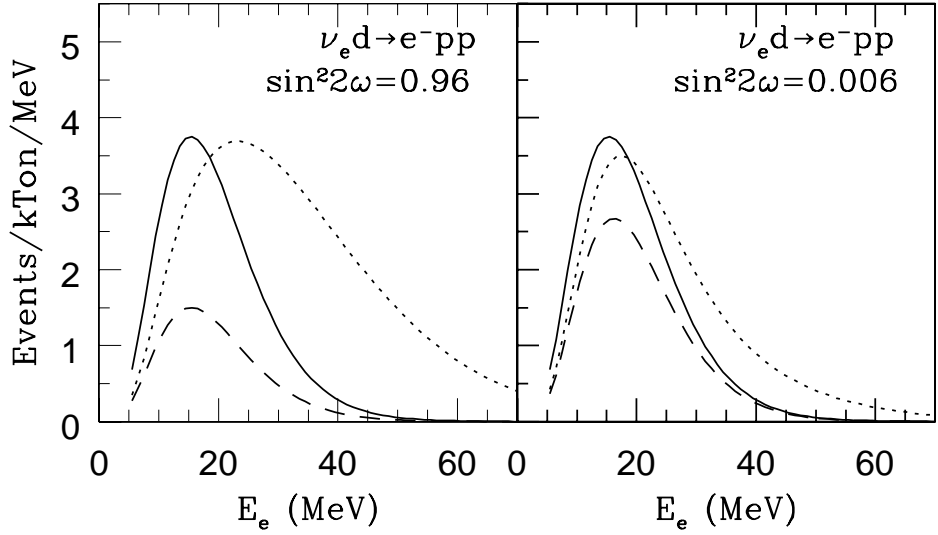}}
\caption{$\nu_e d$ event rates when the upper resonance is completely
non-adiabatic. The solid lines represent the no-mixing case.  The dotted
and dashed lines are due to the effects of 3- and 4-flavour mixing.
Results are shown for two different values of $\omega$ when $\epsilon =
10^{-4}$.}
\label{fig:nuenonad} % fig 5
\end{figure}

\begin{figure}[htp]
\vskip 9truecm
{\includegraphics{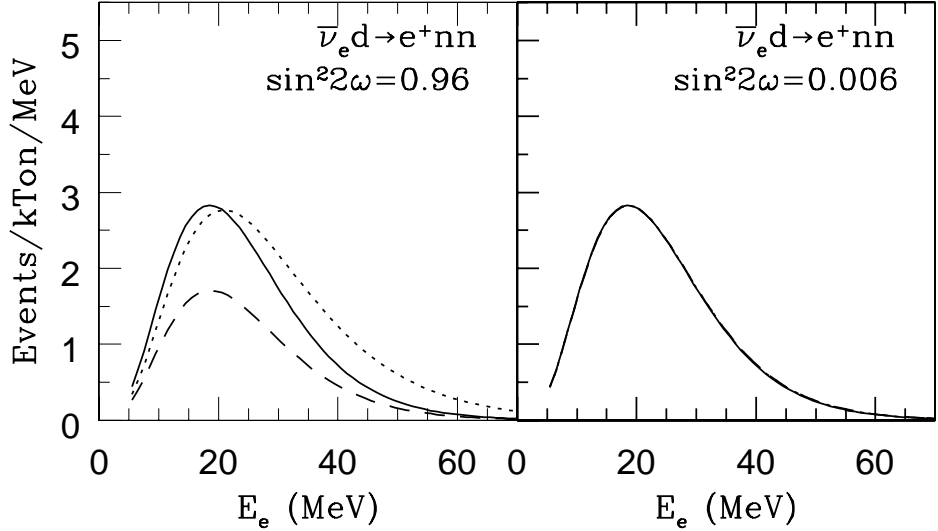}}
\caption{$\overline{\nu}_e\, d$ event rates are shown as a function of
positron energy, $E_e$.  The solid lines represent the no-mixing case.
The dotted and dashed lines are due to the effects of 3- and 4-flavour
mixing. Results for two different values of $\omega$ are shown.}
\label{fig:nuebar} % fig 6

\end{figure}

\begin{figure}[htp]
\vskip 18truecm
{\includegraphics{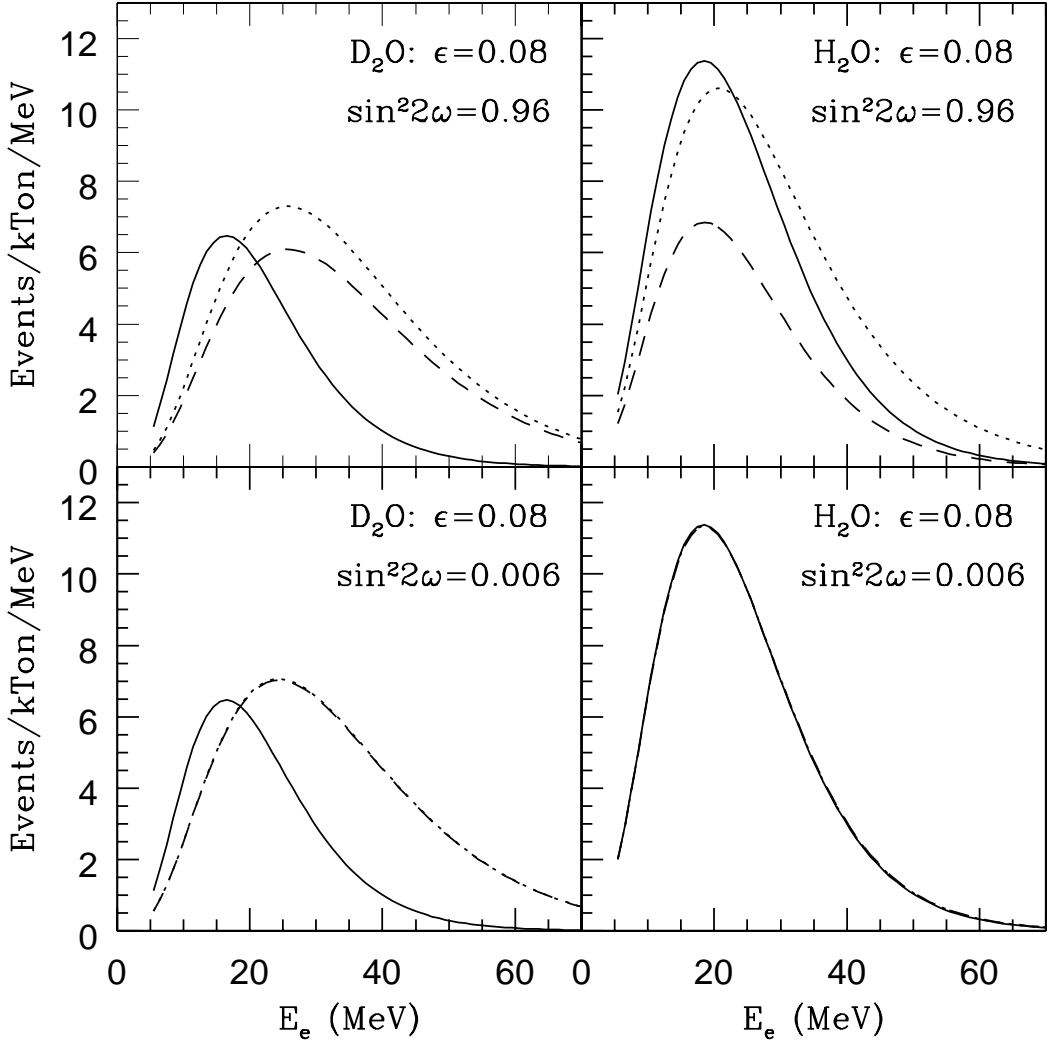}}
\caption{Total event rates (from combining the indivdual $\nu_e\, d$
and $\overline{\nu}_e\, d$ rates shown in Figs.~\ref{fig:nuead} and
\ref{fig:nuebar}) are shown as a function of the electron/positron
energy, $E_e$, for two different values of $\omega$, and for $\epsilon
= 0.08$ so that the propagation is fully adiabatic. The dotted and
dashed lines are due to the effects of 3- and 4-flavour mixing.
Results from a 1 kton water detector (from $\overline{\nu}_e\, p$
alone) are shown on the right, for comparison.}
\label{fig:total} % fig 7
\end{figure}

\begin{figure}[htp]
\vskip 18truecm
{\includegraphics{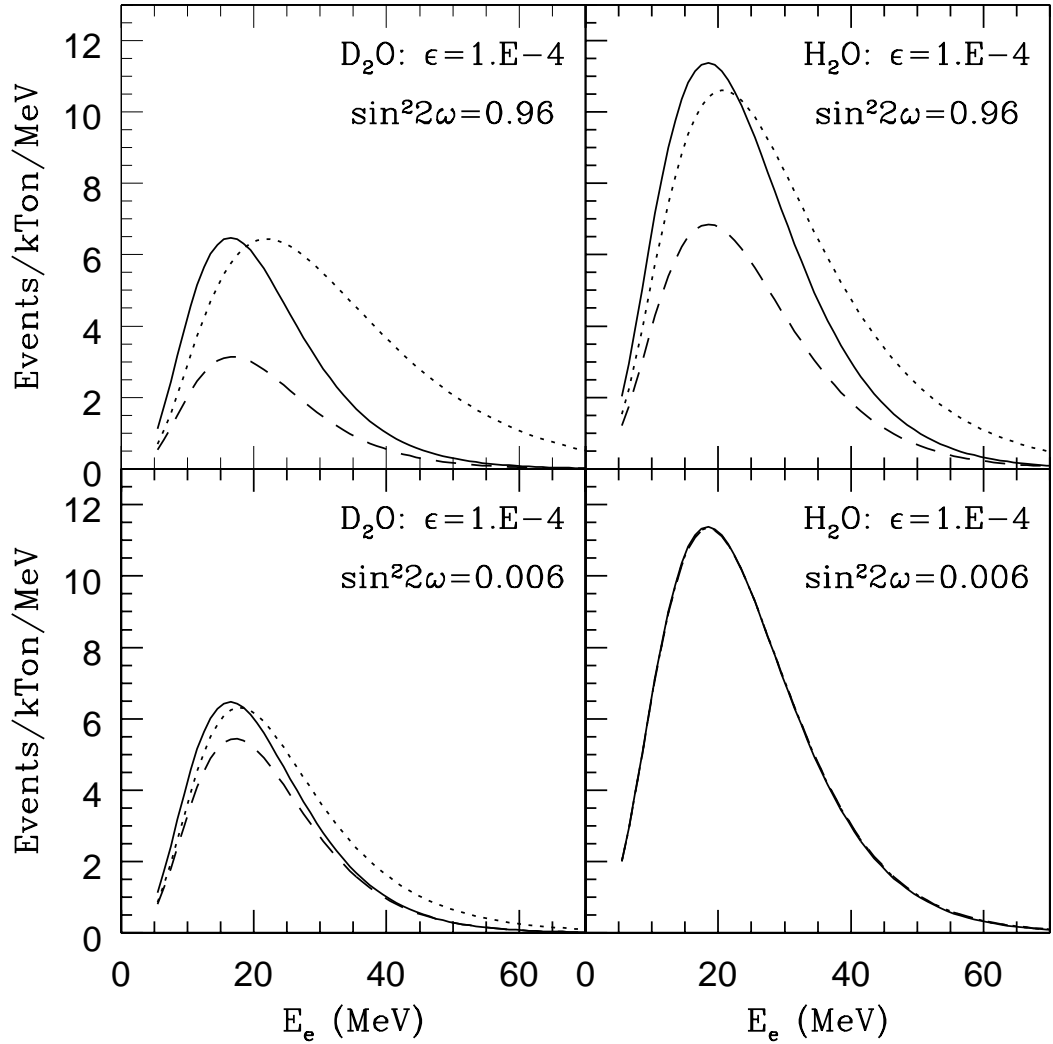}}
\caption{Same as Fig.~\ref{fig:total} but for $\epsilon \sim 0$ so that
non-adiabatic effects are included. Hence this is a combination of
Figs.~\ref{fig:nuenonad} and \ref{fig:nuebar}.}
\label{fig:totaln} % fig 8

\end{figure}

The CC event rates on deuteron computed using the above inputs are
displayed in Figs.~\ref{fig:nuead}--\ref{fig:totaln}. The
time-integrated event rates per unit electron energy bin (of 1 MeV) are
shown as a function of the energy $E_e$ of the detected electron. The
solid lines in all the figures refer to the case when there is no
mixing and serve as a reference. Results for 3- and 4-flavour mixing
are displayed in each of these figures as dotted and dashed lines
respectively.

Fig.~\ref{fig:nuead} shows the predictions for the ($\omega$-independent)
adiabatic $\nu_e\, d$ CC interaction when $\epsilon = 0.08$. Mixing
enhances the high $E_e$ event rates for both 3- and 4-flavour mixing,
which cannot be distinguished here. Furthermore, mixing shifts the peak
of the spectrum to higher energies, from 15 MeV to about 28 MeV because
of the admixture of the hotter $\nu_x$ spectrum and its significantly
higher cross-section with deuterium. This high energy shift should be
clearly observable.

Nonadiabaticity at the upper resonance occurs when $\epsilon \sim 0$.
Then the adiabaticity at the lower resonance is determined by the value
of $\omega$. In fact, $P_L = 0$ unless $\omega$ is small, 
$\sin\omega~{\stackrel{{}_{\displaystyle <}}{{}_{\displaystyle
\sim}}}~0.2$. Hence this is relevant only for the small angle MSW
solution (SMA). We show the dependence of $P_L$ on the neutrino energy
for the small-$\omega$ 3-flavour case in Fig.~\ref{fig:PL}. The
4-flavour result is similar, with a scale factor roughly 0.7. It is seen
that $P_L$ increases with energy,although it still does not reach unity
for the relevant supernova neutrino energies. For larger $\omega$, that
is, for the large angle MSW (LMA) and the vacuum (LMA-V) solutions,
$P_L = 0$.

In Fig.~\ref{fig:nuenonad} we show the electron spectrum for the
non-adiabatic $\nu_e\,d$ CC interaction  when $\epsilon $ is small, in
fact near zero. The fully non-adiabatic case corresponding to small
$\omega$ is shown in comparison with that for a larger value of
$\omega$ in the figure. Here, 3- and 4-flavour mixing give drastically
different results, but the small-$\epsilon$ scenario is in general not
very sensitive to the chosen values of $\omega$.

In Fig.~\ref{fig:nuebar} we show the positron spectrum due to
$\overline\nu_e\,d$ CC interactions for two different choices of
$\omega$. (Here $\epsilon = 0.08$, but the results are essentially the
same even if $\epsilon$ is nearly zero since this sector is always
adiabatic.) Mixing has appreciable effects only for large $\omega$;
however, mixing does not affect the peak position, unlike in the
adiabatic $\nu_e$ case.

We would like to point out that the pure $\nu_e$ spectrum may not be
observable in heavy water. It may be separated out from the total
events sample if the $\overline{\nu}_e$ spectrum can be reliably
separated out by various detection techniques such as looking for two
neutrons in coincidence with the positron\footnote{We thank the referee
for pointing out this possibility.}. However, we will
show that the total number of events will still be sensitive to
the mixing parameters. We will therefore discuss both the total as well
as the individual $\nu_e$ and $\overline{\nu}_e$ spectra. 

The total event rates due to the dominant CC $\nu_e\,d$ and
$\overline{\nu}_e\,d$ processes are shown in Figs.~\ref{fig:total} and
\ref{fig:totaln} for the two choices of $\epsilon$. The total isotropic
CC events in water, due to $\overline{\nu}_e\,p$ alone, are also shown
for comparison. The most significant difference between the two is that
of the adiabatic case with small $\omega$ (the lower two panels of
Fig.~\ref{fig:total}) which is independent of the number of flavours.
This is because of the enhancement in the $\nu_e$ events, independent
of $\omega$. At all $\omega$, the peak is at a higher energy than
expected from the no-mixing case in heavy water but remains the same
for a water detector. This shift may be sufficiently significant and
therefore observable, in the adiabatic scenario, for all $\omega$.
Finally, the upper two panels of Fig~\ref{fig:total} indicate that a
significant depletion in the observed events in water,
together with an enhanced number of events in heavy water is an
unambiguous signal of 4-flavour mixing with large $\omega$ (dashed lines
in Fig.~\ref{fig:total}).

The corresponding results for the dominant CC events in the non-adiabatic
case are shown in Fig.~\ref{fig:totaln}. Here there is no significant
shift in the spectral peak. Also, the signals in water and heavy water
are very similar, with the signals being either enhanced or depleted
similarly in both. For instance, the large $\omega$ 4-flavour
signal shows depletion both in water and heavy water. This may be
difficult to distinguish from the no-mixing case if the overall
normalisation of the supernova spectrum is uncertain by more than a
factor of two. In all cases, the small-$\epsilon$, small-$\omega$
scenario also cannot be distinguished from the no-mixing case. Hence the
small-$\epsilon$ case may be difficult to establish unambiguously,
independently of the supernova model inputs.

Keeping in mind that the supernova dynamics may have large uncertainties,
we will later also analyse the ratios of the total number of events in
water and heavy water detectors. These are likely to be less sensitive
to the supernova models (although they do depend on the temperature
hierarchy for $\nu_e$, $\overline{\nu}_e$ and $\nu_x$) and hence may be
more robust signals of mixing.

\subsection{Integrated number of events}
The predicted time integrated number of events resulting in a scattered
electron with energy, $E_e > 5$ MeV (which is a typical threshold for
Cerenkov detectors), are shown in Tables~\ref{tab:CCad} and
\ref{tab:CCnonad}, for the adiabatic and non-adiabatic cases
respectively. As before, the number is calculated assuming a supernova
explosion at 10 kpc for a 1 kton detector. Listed are the dominant
CC events on deuterons in heavy water and free protons in
water, along with the elastic scattering events on electrons and the CC
events on oxygen nuclei.

For heavy water, we have listed the individual contributions from
$\nu_e$ and $\overline{\nu}_e$ on deuteron. In water, the corresponding
dominant events are from $\overline{\nu}_e$ on $p$. The elastic events
are from $\nu_e\, e$, $\overline{\nu}_e\, e$, $\nu_x\, e$, and
$\overline{\nu}_x\, e$. Since they will all be detected in the extreme
forward direction, they have been summed up and listed as total
$\nu \, e$ events in the tables. The oxygen events, listed as $\nu\,O$,
include both $\nu_e\, O$ and $\overline{\nu}_e\,O$ CC events, which will
predominantly be in the backward direction, especially when enhanced by
mixing. In particular we
tabulate the events for 3- and 4-flavour mixing when $\omega$ is both
small and large.

In Table~\ref{tab:CCad}, we show the results for the fully adiabatic
case when $\epsilon=0.08$.  It is seen that the bulk of the events
(more than 90\%) are from the dominant events on $p$ or $d$.  Mixing
always enhances the $\nu_e\,d$ channel by more than a factor of two;
hence adiabatic propagation always predicts an enhanced rate of total
events in heavy water even though there is reduction in the
$\overline{\nu}_e\,d$ channel (the other dominant process) for some
parameters. In contrast, the total number of events in water may even
go down as compared to the no-mixing case, depending on the parameter
values.

In Table~\ref{tab:CCnonad}, we show similar predictions for the case
when $\epsilon \sim 0$ or when the upper resonance(s) become fully
non-adiabatic. The value of $\omega$ then determines whether or not the
lower resonance is adiabatic. Recall that the antineutrino propagation
is always adiabatic. Again, the contribution from the $\nu \, e$ and
$\nu \, O$ events is small compared to the dominant events. Small
changes in the $\overline{\nu}_e\,p$ and $\overline{\nu}_e\,d$ events
between Tables~\ref{tab:CCad} and \ref{tab:CCnonad} are due to changes
in the value of $\epsilon$.  Irrespective of the parameter values, it
is seen that there is never any depletion in the 3-flavour case. As in
Fig.~\ref{fig:totaln} an interesting scenario occurs when $\omega$ is
large in the 4-flavour case. Here the total number of dominant events in
both $\nu_e\, d$ and $\overline{\nu}_e\,d$ in heavy water and in
$\overline{\nu}_e\, p$ in water, is reduced by a factor proportional to
$\cos^2\omega$. This is the only scenario where there is depletion in
both water and heavy water.

\subsection{Possible discrimination of various ``mixing models''}

So far, we have assumed that most of the mixing parameters are known
and used the supernova measurement as a potential check for
self-consistency of the model parameters. This is because there is
still very little known about the supernova neutrino spectrum through
observations and hence there is both theoretical and experimental
uncertainty about the details of the neutrino spectrum. However, it is
still instructive to actually turn the question around and ask, suppose
another supernova explosion is observed through its neutrino emission.
Will an excess over (or a depletion from) the expected number of events
unambiguously determine some of the model mixing parameters ? The
answer to this question can be obtained from Table~\ref{tab:r}. It is
seen that certain classes of models may be ruled out, depending on the
observation.  We define the ratio of the total number of events (from
all possible interactions with an electron (or positron) in the final
state) potentially observed from a future supernova (equivalently, the
prediction from a given neutrino mixing model) to the expected number
of events without mixing (for $E_e > 5$ MeV):
\begin{equation}
R_i = \by{\mbox{\rm observed number of events}}
{\mbox{\rm calculated number without mixing}}~,
\label{eq:Ri}
\end{equation}
where $i=D,H$ refer to 1 kton heavy water and water detectors
respectively.  The denominator refers to the the expected number of
events (using a standard supernova model, with no mixing) as computed
from a Monte Carlo simulation that
takes into account detector resolution, efficiency, etc., consistent
with the detector at which the events were observed. An observation may
find $R_i > 1$, $R_i < 1$ or $R_i \sim 1$. Note that even though the
ratio refers to the total number of events, the inferences drawn reflect
mainly the behaviour expected from the dominant CC processes on protons
in water and deuterons in heavy water. The mixing models (with
model parameters $\epsilon$ and $\omega$, including adiabaticity)
consistent with, or predicting, such an observation are shown in
Table~\ref{tab:r}.  Here the non-adiabaticity, that is, the value of
$P_L$ at the lower resonance has been computed assuming a typical
value of $\delta_{21} = 10^{-5}$ eV${}^2$. Water detectors cannot
distinguish adiabatic (A) and non-adiabatic (N) scenarios, that is,
whether or not $\epsilon$ is different from zero, but can distinguish
the number of flavours when $\omega$ is large (see the last column of
Table~\ref{tab:r}). In $D_2O$, however, most models predict $R_i > 1$.
The {\it only} observation of $R_i < 1$ in $D_2O$ occurs for the 4-flavour
non-adiabatic case with $\epsilon \sim 0$ and large $\omega$.

On combining data from water and heavy water detectors, an improved
discrimination of model parameters is possible, as can be seen from
Table~\ref{tab:rdh}. Here the different values of $R_H$ and $R_D$ are
listed, along with the models that are consistent with such a
combined observation. First of all, it is seen that combining the two
measurements immediately allows for a separation between the
adiabatic and non-adiabatic cases and hence whether $\epsilon$ is
different from zero, except in the 3-flavour case with large $\omega$.
It should be noted that the $\epsilon, \omega \to 0$ scenario is
unlikely to be distinguished from the no-mixing case. It is also seen
that certain combinations of $R_D$ and $R_H$ do not occur for any of
the allowed parameters values. For instance a depletion in heavy water
is {\it only} consistent with depletion in water as well. Any other
result in water indicates that the overall normalisation of the
spectrum is probably in error. An occurrence of such ``forbidden''
combinations may therefore can be used as a check on the overall
normalisation of the supernova spectrum. This result of course is
limited to the class of models we are analysing here. 

The following scenario is best
suited to determining the value of $\omega$. (1) There are fewer
isotropic events than expected ($R_H < 1$) in a water Cerenkov detector
such as SuperK. This reduction factor determines $\cos^2\omega$. (2)
The {\it same} reduction factor ($R_D < 1$) fits the data from a heavy
water detector such as {\sc sno}. This can imply that the correct
mixing matrix is one with 4 flavours and large $\omega$, with
non-adiabatic neutrino propagation. (3) If on the other hand there are
enhanced number of events ($R_D > 1$) at the heavy water detector, it
clearly indicates adiabatic neutrino propagation. This in turn implies
that $\epsilon$ is different from zero which has so far been claimed
only by {\sc lsnd}.  Variation of supernova input parameters, which we
will discuss in the next section, does not alter this result.

These qualitative features can be quantified by defining the double
ratio,
\begin{equation}
R_{D/H} = \by{R_D}{R_H}~,
\end{equation}
which is independent of the overall normalisation of the neutrino flux
and hence provides a better diagnostic. In practice, it may not be
possible to directly take a ratio of the data from water and heavy
water detectors since the two measurements will differ in their
systematics, apart from such considerations as detector efficiency and
resolution. Since $R_i, i= D,H$ are normalised to the theoretical
expectancy including these considerations, $R_{D/H}$ is not likely to
be sensitive to details of detector design and can thus provide a
robust, quantitative indicator of different types of mixing.

This double ratio (where $R_i$ has been calculated as before) has been
shown in Figs.~\ref{fig:newr} and \ref{fig:newrc} as a function of
$\omega$. The ratio is plotted for the total number of events with the
cut on the observed electron (positron) energy, $E_e \ge 5$ MeV in
Fig.~\ref{fig:newr}. The two curves in each figure correspond to the
two values of $\epsilon$ when the propagation at the upper resonance(s)
is purely adiabatic, $\epsilon = 0.08$ (solid lines) or purely
non-adiabatic, when $\epsilon = 10^{-4}$ (dashed lines). These are the
two cases that a water detector is normally not able to resolve.  Also
shown are dotted vertical lines corresponding to the solutions allowed
by the solar neutrino problem:  $\sin^2 2\omega = 0.006, 0.76$ and
$0.96$. Non-adiabaticity at the lower resonance has been computed
using $\delta_{21} = 10^{-5}$ eV${}^2$ as before.  While the 3-flavour
mixing case is shown on the left, the 4-flavour result is plotted on
the right.

Obviously, a value of unity is expected for the case of no-mixing. We
analyse each case in turn.
\begin{enumerate}
\item
We see from Fig.~\ref{fig:newr} that the double ratio $R_{D/H}$ is
always strictly greater than one for the adiabatic case, independent of
$\omega$ or the number of flavours, $f$.
\item
Even in the non-adiabatic case, it can be less than one only when $f=4$.
Note however that currently allowed values of $\omega$ lead to $R_{D/H}
\sim 1$ in the non-adiabatic case. The case $R_{D/H} < 1$ occurs only
for intermediate values of $\omega$.
\item 
For $f=3$ (left panel of Fig.~\ref{fig:newr}), the double ratio at
small $\omega$ is different for the adiabatic and non-adiabatic cases,
which may therefore be distinguished. However, it may be very
difficult to distinguish these two for large values of $\omega$.
\item
On the other hand, these two cases are easily distinguished for all
$\omega$ for $f=4$ as can be seen from the right panel in the
figure.
\item
While the double ratio is similar for both $f=3,4$ for small $\omega$,
$\sin^22\omega \le 0.1$, the number of flavours can be distinguished
for larger values of $\omega$, especially in the currently allowed
region, only in the adiabatic case.
\item
However, as stated before, independent of $f$, the small $\omega$
non-adiabatic solution with $\epsilon \sim 0$ cannot be distinguished
from the no-mixing case.
\end{enumerate}
Keeping in mind that a thermal neutrino flux distribution such
as the one we have used may overestimate the high energy spectrum, the
case $5 \le E_e$ (MeV) $\le 40$ is shown separately in
Fig.~\ref{fig:newrc}. Most of the features survive the cuts; hence this
ratio is likely to be a stable indicator of mixing.

Finally, we note that the denominator of the double ratio is dominated
by $\overline{\nu}_e$ events. Hence, the {\it same} discriminatory
power can be achieved using data from a heavy water detector {\it
alone} if it is possible to separate the $\nu_e\,d$ and
$\overline{\nu}_e\,d$ events.  As stated earlier, this may be possible,
for instance, at SNO, by detecting both the neutrons in coincidence
with the positron emitted in the $\overline{\nu}_e\,d$ interaction. SNO
is also planning to increase the neutron detection efficiency (to more
than 80\%) by adding salt to the heavy water \cite{SNOback}. In this
case, the double ratio, defined for heavy water alone,
\begin{equation}
r_d = \by{R_{e+\overline{e}}}{R_{\overline{e}}}~,
\end{equation}
will provide as much information as the double ratio $R_{D/H}$. Here, 
$$
R_{\overline{e}} = \by{\hbox{observed number of }\overline{\nu}_e
\hbox{ events}} {\hbox{calculated number without mixing}}~,
$$
while $R_{e+\overline{e}}$ is a similar ratio, defined for the total
number of events from both $\nu_e$ and $\overline{\nu}_e$ interaction
with deuterium:
$$
R_{e+\overline{e}} = \by{\hbox{observed number of }\nu_e
\hbox{ and } \overline{\nu}_e \hbox{ events}}
{\hbox{calculated number without mixing}}~.
$$
Both $R_e$ and $R_{e+\overline{e}}$ are calculated for a heavy water
detector. To a very good approximation, we have
\begin{equation}
R_{D/H} \approx r_d~,
\label{eq:sumrule}
\end{equation}
The approximation arises partly from ignoring events due to electron
and oxygen targets in $r_d$. The error also arises from the differences
in the denominators of the two ratios, one involving
$\overline{\nu}_e\,p$ and the other $\overline{\nu}_e\,d$.  Despite a
mild energy dependence of the ratio of these two cross-sections
\cite{crossd,kubo}, it turns out that the ratio of the total events
expected from these two processes remains in the range $4 \pm 0.1$ (see
Table~\ref{tab:CCad}). This is true both when there is no mixing, and
with mixing, for all allowed values of mixing angles $(\phi, \omega)$.
Hence this factor cancels when the ratio $R_{\overline{e}}$ is
expressed in terms of $R_H$.  Hence the approximation in
Eq.~\ref{eq:sumrule} should be valid to within a few percent. In
addition, the double ratio $r_d$ will have the advantage of reduced
systematic errors, since data from different experiments do not have to
be combined in order to calculate it. Hence it will be useful to
calculate such a ratio, by separating out the $\overline{\nu}_e$ events
on deuteron.

\subsection{Sensitivity to supernova model parameters}

So far, we have discussed the sensitivity of the supernova neutrino
spectrum to various neutrino mixing parameters. However, the supernova
model parameters (temperature and luminosity) are themselves uncertain
and still need to be experimentally established. It is therefore
important to study the effect of variation of these parameters on the
results we have so far obtained.

Supernova dynamics is a very complicated issue. Here we will follow a
simple-minded approach. Changes in the luminosity affect the overall
normalisation while changes in the temperature (or average energy)
change in the {\it shape} of the spectrum. Variations in these
parameters, while being time-dependent, are not random, but
systematic.  For instance, the supernova model parameters depend on the
protoneutron star mass (an increase of which increases both the average
energy and luminosity of neutrinos) as well as the underlying
high-density equation of state and the initial conditions. The effect
of this on the total neutrino spectrum has been studied in
Ref.~\cite{SNerrors}. (The temperature variation in the spectra of
individual flavours is not discussed). The study indicates that the
typical temperature variation of the total neutrino spectrum at all
times does not exceed about $\pm 1$ MeV. While the variation due to
uncertainties in the initial conditions is relatively small, the
average energy systematically decreases for smaller protoneutron mass
stars and those evolving with a stiffer equation of state. On the other
hand, the luminosities are virtually identical for all these cases
until a time $t \sim 10$ secs., by which time most of the detectable
neutrinos are emitted.

We will first estimate the systematic errors due to uncertainties in
the supernova temperature. In the absence of detailed information on
temperature variation of the individual flavours, we shall assume a
systematic time-independent increase (or decrease) in the temperature
of spectra of all flavours by 1 MeV. This will then be an estimator of
the outer limits of variation of the results from the original
calculation.

\begin{figure}[htp]
\vskip 9truecm
{\includegraphics{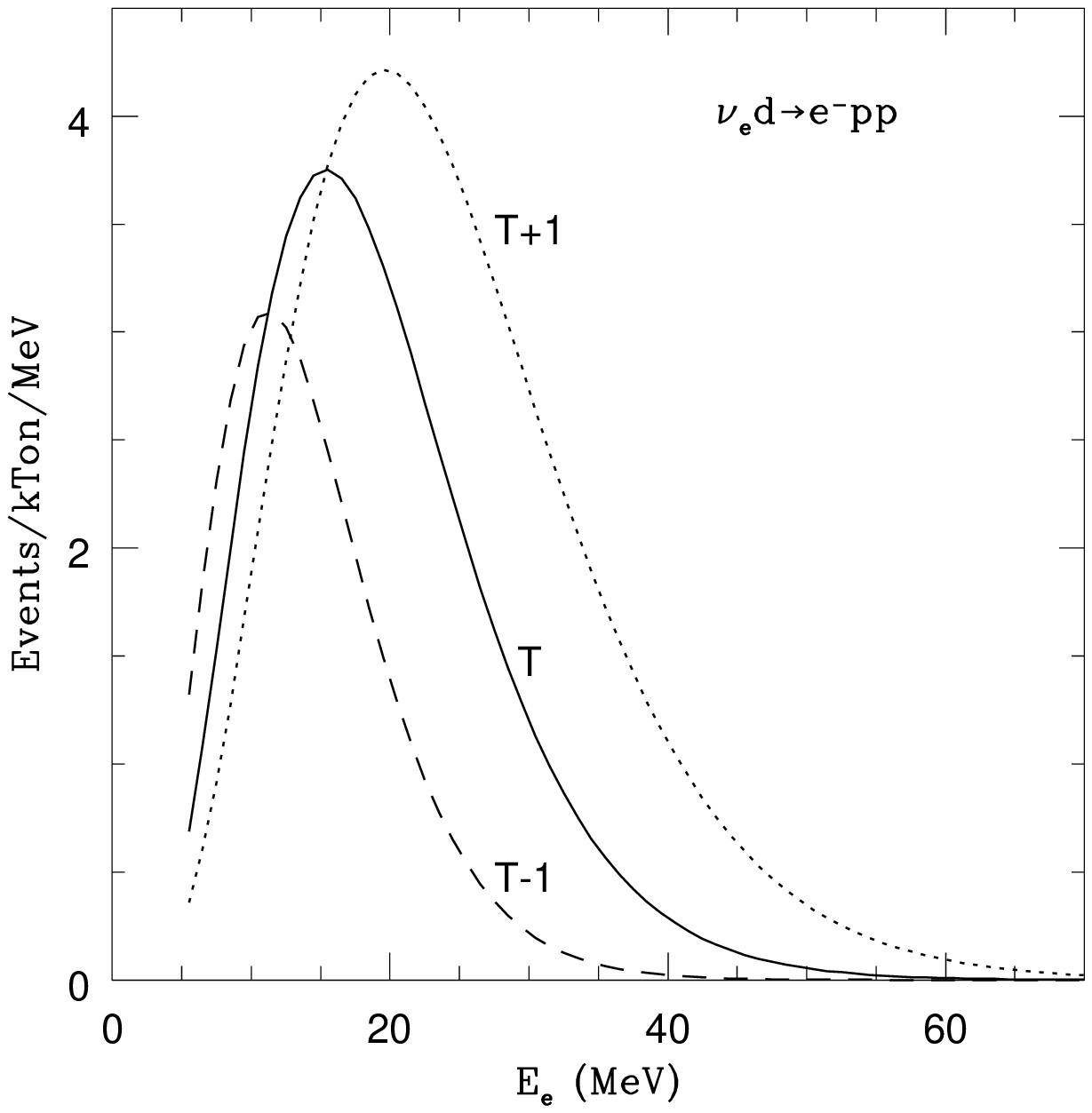}}
\caption{The same as Fig.~\ref{fig:nuead} for electron neutrinos in the
no-mixing case (solid lines). Added are the dotted and dashed lines
corresponding to the case when the supernova neutrino spectral
temperatures are increased or decreased by 1 MeV in each time-bin.}
\label{fig:nueerr} % fig 11
\end{figure}

Fig.~\ref{fig:nueerr} shows the expected number of events due to
$\nu_e\,d$ interaction in the absence of mixing and when the temperature
is systematically increased or decreased by 1 MeV in all time bins,
$$
T_b(\alpha) \to T_b(\alpha) \pm 1 \hbox{ MeV}~.
$$
The base-line supernova spectrum is shown in comparison as a solid
line.  There is of course a shift in the spectral peak (by around 3 or
4 MeV); however, there is a large change in the high energy part of the
spectrum (accentuated due to the $E^2$ dependence of the
cross-section). It will still be possible to distinguish the adiabatic
mixing case since the increase at high energy in this case is
substantially larger than from errors in the supernova spectrum.
However, other cases, especially the non-adiabatic cases, will not be
clearly distinguishable. It must be noted that in any event the
spectral peak for $\nu_e$ events is a good index of the temperature of
the spectrum, either of the unmixed $\nu_e$ or of the hot $\nu_x$
spectrum.

\begin{figure}[htp]
\vskip 9truecm
{\includegraphics{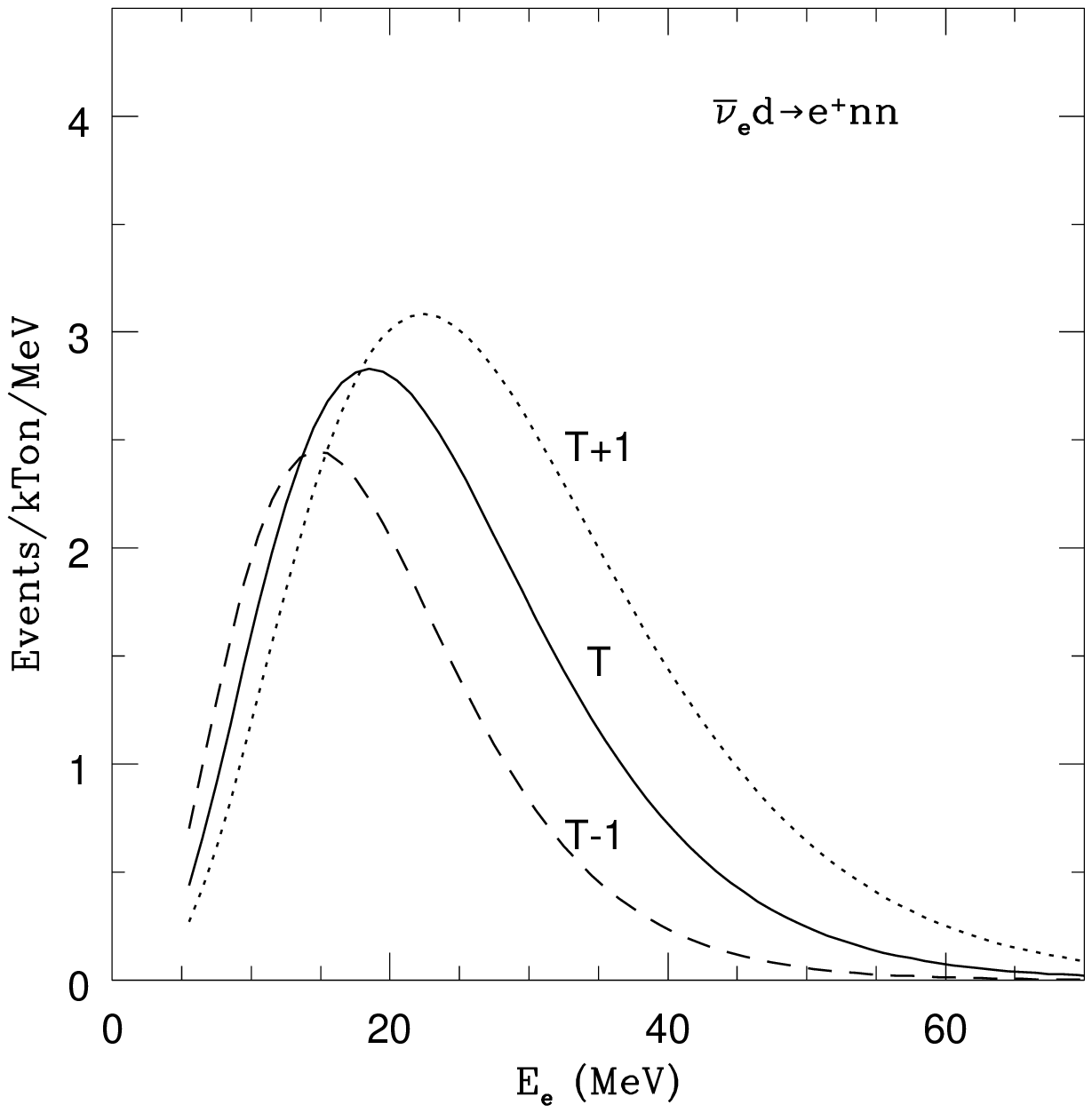}}
\caption{The same as Fig.~\ref{fig:nuebar} for electron antineutrinos
in the no-mixing case (solid lines). Added are the dotted and dashed
lines corresponding to the case when the supernova neutrino spectral
temperatures are increased or decreased by 1 MeV in each time-bin.}
\label{fig:nuebarerr} % fig 12
\end{figure}

Fig.~\ref{fig:nuebarerr} shows the results for the case of the
$\overline{\nu}_e\,d$ unmixed spectrum. There is a similar dependence
(since the energy dependence of the cross-section is the same as in the
$\nu_e\,d$ case. Since mixing does not significantly shift the
$\overline{\nu}_e$ spectral peak (see Fig.~\ref{fig:nuebar}), this will
remain a good indicator of the corresponding spectral temperature.

While there are large variations in the results for the individual
spectra, these will be cancelled out in the double ratio $R_{D/H}$ of
the events in heavy water and water (in fact, this is the purpose of
constructing such a ratio). This can be seen from
Fig.~\ref{fig:newrerr}, where the same double ratio, $R_{D/H}$, is
plotted for the different temperature sets, $T$ (as solid lines) and
$T\pm 1$ (as dashed lines). It is likely that any time-dependence of
the temperature variation that we have ignored, will affect the
numerator and denominator of the ratio in the same way; hence inclusion
of time dependence should not affect this analysis. In computing this
ratio, the ``observed number of events'' as required for the
calculation of $R_i$, $i = D,H$, in Eq.~\ref{eq:Ri} is now determined
both by the mixing parameters as well as by the modified supernova
model. It is seen that there is very little sensitivity to the
variations in the temperature.  This is especially so in the adiabatic
case. A greater sensitivity to the model parameters in the small
$\omega$, small $\epsilon$ (non-adiabatic) case occurs because of the
presence of the additional energy-dependent factor, the Landau-Zener
transition probability, $P_L$. Hence inclusion of temperature
variations in the supernova model does not change the conclusions about
discrimination of different mixing models.

We add a note on variations in the luminosity due to, for example,
choice of different initial conditions \cite{SNerrors}. This affects
the overall normalisation which is an indicator of the total energy
emitted.  However, the double ratio will not be sensitive to this,
unless these changes are extremely time-dependent. It is doubtful
whether reasonable conclusions can be drawn in such a case, unless
there are significantly large numbers of events in each time bin. In
short, while the individual flavour spectra can be significantly
modified by uncertainties in the supernova model parameters, the
double ratio $R_{D/H}$ is largely insensitive to such variations. Hence
it is a good indicator of mixing.

A final remark about statistical errors. As already stated, the
background to a supernova signal due to both radioactivity as well as
solar events is small at SuperKamiokande and SNO. Hence the signal will
be clearly defined. In any case, the statistical errors (assuming a
$1/\sqrt{N}$ error for both the numerator and the denominator and
adding suitably in quadrature) have been calculated for the double
ratio $R_{D/H}$. This has also been shown in Fig.~\ref{fig:newrerr}.
The errors are so small that they are not visible, except as a slight
thickening of the lines near the $\omega \to 1$ region. Of course, if
the supernova is 50 kpc and not 10 kpc away, the statistical errors
become 5 times larger. Recall however that we have computed the events
in 1 kton of detector. Larger detector volumes will further reduce this
error. In general, the statistical quality of the signal, while being
good, will depend on the size of the detector as well as the distance
to the supernova.

\subsection{Neutral Current events}

As is well-known, heavy water detectors can
directly observe NC events. This is very important in the context of
supernova neutrinos since neutrino emission from supernovae is
practically the only observable system where neutrinos (and
antineutrinos) of all flavours are emitted in roughly equal
proportions. Note that there are also NC events on oxygen targets in
both water and heavy water, with a characteristic signal of photons with
energies in the range of 5--10 MeV \cite{langanke}. However, these
events are fewer in number than the NC events on the deuteron that we
will discuss here.

While there is no loss of NC events in the case of 3-flavour mixing,
the existence of a fourth flavour will be signalled by loss of NC
events into this sterile channel. This can be seen from
Tables~\ref{tab:NCad} and \ref{tab:NCnonad} where the total number of
NC events from neutrinos or antineutrinos of all flavours (with $E_\nu
> 3$ MeV) is listed for different possible values of $\omega$
consistent with the solar neutrino expectation. It is seen that the
number of NC events is not very sensitive to the value of $\omega$;
however, from Table~\ref{tab:NCad} it is clear that in the adiabatic
case, there is about 25\% depletion with 4-flavour mixing when compared
to the no-mixing case. If the value of $\epsilon$ and the overall
normalisation of the spectrum is known, NC current events can be used
to discriminate between three and four flavour mixing. Recall that the
CC events are always enhanced by a factor of 1.5--2 for the adiabatic
case. Hence more conservatively, the NC events can be used to normalise
the supernova spectrum to at least within 25\%.

When the upper resonance is non-adiabatic, however, part of the signal
gets regenerated, especially for smaller $\omega$.  Hence, as
Table~\ref{tab:NCnonad} shows, there are roughly the same number of NC
events with and without mixing in the non-adiabatic case, independent
of the number of flavours. The corresponding CC channel shows severe
depletion only when $\omega$ is large; otherwise, it is either
enhanced, or the same as the no-mixing case.  Hence here also the NC
events can be used to normalise the supernova spectrum.

\section{4-flavour $(3+1)$ mixing scheme}

So far, all results in the 4-flavour analysis referred to the $(2+2)$
scheme as shown in Fig.~\ref{fig:level}b. We briefly discuss results in
the $(3+1)$ scheme where the mixing matrix is defined through: 
\begin{equation}
\left[ \nu_e \quad \nu_\mu \quad \nu_\tau \quad \nu_s \right ]^T = U \times 
\left[ \nu_1 \quad \nu_2 \quad \nu_3 \quad \nu_4 \right ]^T~,
\end{equation}
and
\begin{eqnarray}
U & = & U_{34} \times U_{24} \times U_{23} \times
U_{14} \times U_{13} \times U_{12}~, \nonumber \\
  & = & U_\rho \times U_\epsilon \times U_\psi \times 
  U_\epsilon \times U_\epsilon \times U_\omega~.
\end{eqnarray}
Here the (13) and (14) angles are constrained by {\sc chooz} to be
small: $\theta_{13},\theta_{14} \sim \epsilon \le \epsilon_0$
\cite{dutta2} as in the $(2+2)$ scheme. The atmospheric neutrino
problem now constrains $\theta_{23}$ (the equivalent of the angle
$\psi$ in the 3-flavour case) to be maximal, when $\theta_{24}$ becomes
small, $\theta_{24} \le \epsilon_0$. However, the (34) mixing angle $\rho$
is not constrained by any known experimental data.

Since $\nu_e$ is produced in the supernova core in essentially the
$\nu_4$ mass eigenstate, any non-adiabaticity results in jumps near the
upper MSW resonances. The adiabaticity parameter here will be determined
by the $\theta_{14}$ angle, which is again small, $\theta_{14} =
\epsilon$. However, the adiabaticity parameter at the lower resonance
depends on the unknown angle $\rho$ and hence we do not comment
on the non-adiabatic case here.

The expressions for the fluxes as observed on earth are given in
Eqs.~(\ref{eq:fe}), (\ref{eq:febar}) and (\ref{eq:fx}) with
probabilities $P_{\alpha\beta}$ computed for the $(3+1)$ scheme. The
relevant probabilities, $P_{ee}$ and $P_{es}$, for the CC events in the
adiabatic case are {\it independent} of the unknown angle $\rho$ and
are in fact the same in both the $(2+2)$ and $(3+1)$ schemes. Hence all
the results (as shown in Figs.~\ref{fig:nuead}, \ref{fig:nuebar},
\ref{fig:total}) for the CC events on deuteron in the adiabatic sector
are insensitive to the position of the sterile neutrino in the case of
4-flavours. This is true for the CC events on oxygen as well.

In the NC case, we need the probabilities $P_{se}$ and $P_{ss}$ which
are different from the $(2+2)$ case. Ignoring small terms of order
${\cal O} (\epsilon)$, we have,
\begin{eqnarray}
P_{se} & = & c_\rho^2~, \nonumber \\
P_{ss} & = & s_\rho^2/2 ~, \nonumber \\
P_{\overline{s}\overline{e}} & = & s_\omega^2 s_\rho^2/2 ~, \nonumber \\
P_{\overline{s}\overline{s}} & = & c_\omega^2 s_\rho^2/2 ~, 
\end{eqnarray}
where $c_\rho, s_\rho$ are $\cos \rho$ and $\sin\rho$ respectively.

While these fluxes do depend on $\rho$ it turns out that the
suppression factor in the adiabatic case is again around 75\% for both
large and small values of $\omega$, again independent of the value of
$\rho$. This is because the dominant contribution to the NC
sector is from the $F_x$ and $F_{\overline{x}}$ fluxes, as in the
$(2+2)$ case. These terms are very weakly dependent on the unknown
angle $\rho$. Hence in the NC sector as well, the (adiabatic)
$(3+1)$ 4-flavour scheme gives almost the same predictions as the $(2+2)$
scheme. Because of this, there will not be much difference in the
elastic events on electrons as well.

\section{Summary and Discussion}

To summarise, we have contrasted
signals from supernova neutrinos (antineutrinos) in water and heavy
water detectors. We include the dominant charged current events from
deuteron targets in heavy water, and proton targets in water, as well as
the elastic scattering off electrons and charged current events on
oxygen in both detectors. In all cases, an electron (or positron) is
detected in the final state. The detailed distribution of events as a
function of the scattered electron (positron) energy depends on the
number of flavours and the mixing parameters in a complex manner. We
have discussed all these cases.

\begin{figure}[htp]
\vskip 8truecm
{\includegraphics{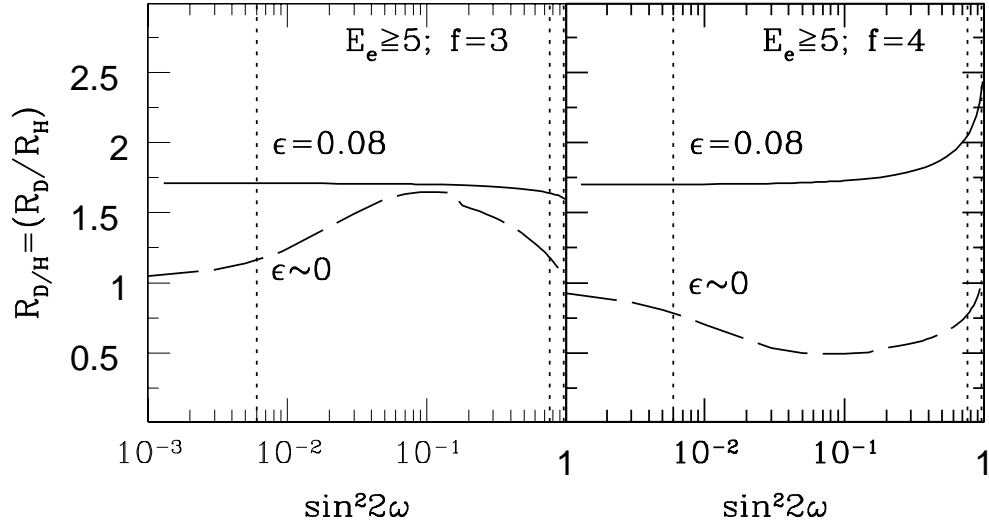}}
\caption{The double ratio $R_{D/H}$ of the ratio of the total events
observed through the detection of an electron (or positron) with
$E_e \ge 5$ MeV, from a future supernova explosion to
that expected, in a heavy water and a water detector, shown as a
function of the (12) mixing angle $\omega$. Solid and dashed lines
correspond to adiabatic ($\epsilon = 0.08$) and non-adiabatic
($\epsilon \sim 0$) neutrino propagation at the upper resonance(s). The
case for 3-flavour mixing is shown on the left and that for 4-flavours
on the right. The vertical dotted lines indicate the currently
favoured values of $\sin^22\omega$ according to solar neutrino
analysis.}
\label{fig:newr} % fig 9
\end{figure}

\begin{figure}[htp]
\vskip 9truecm
{\includegraphics{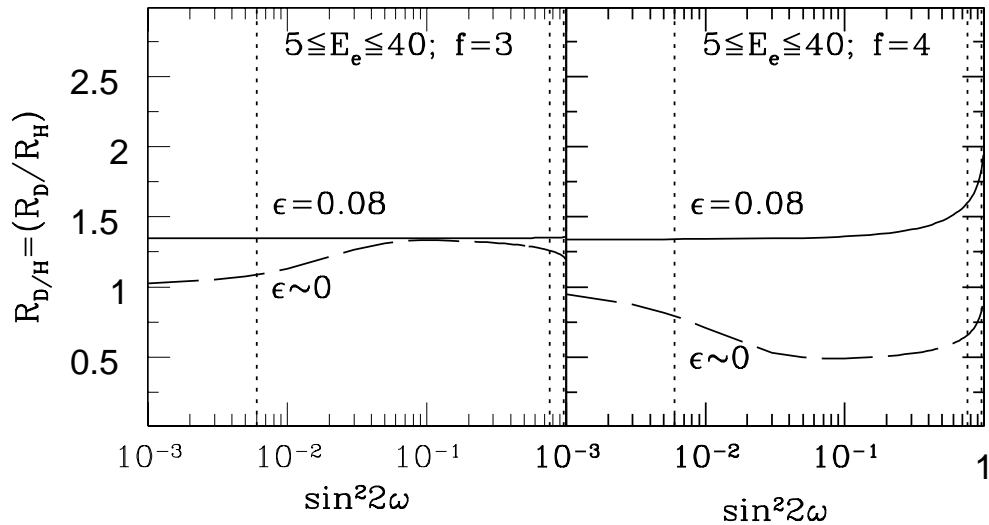}}
\caption{The same as fig.~\ref{fig:newr}, with a high energy cut on the
electrons, $5 < E_e < 40$ MeV, to decrease sensitivity to the
high energy tail of the neutrino spectrum.}
\label{fig:newrc} % fig 10
\end{figure}

\begin{figure}[htp]
\vskip 9truecm
{\includegraphics{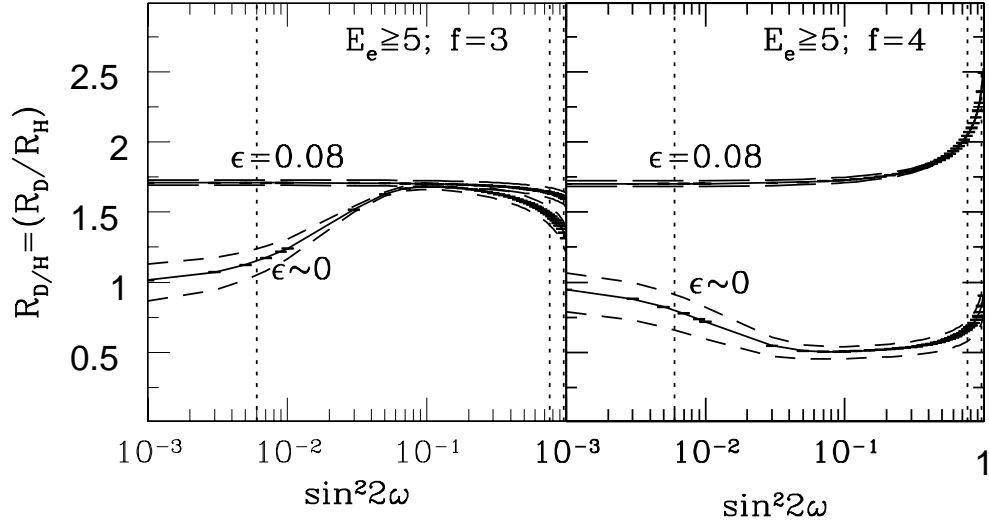}}
\caption{The same as fig.~\ref{fig:newr}, but where the dashed curves
now indicate the changes in the double ratio $R_{D/H}$ due to variations
in temperature in the supernova model by $\pm 1$ MeV. The statistical
errors are also plotted but are too small to be distinguished. }
\label{fig:newrerr} % fig 13
\end{figure}

In particular, Figs.~\ref{fig:newr} and \ref{fig:newrc} show the
combined sensitivity of water and heavy water detectors to the neutrino
mixing parameters for such events by defining a double ratio of the
observed to expected number of events in heavy water and water
detectors respectively. These results reflect essentially the behaviour
of the dominant charged current $\nu_e$ and $\overline{\nu}_e$ events
on deuterons which are comparable to the $\overline{\nu}_e$ charged
current interaction on protons in water. However, its dependence on the
mixing parameters is very different from that for water. It turns out
that a comparison of the signals from water and heavy water detectors
can yield important information on not only mixing parameters but also
on the number of flavours involved. While 3-flavour mixing typically
results in an enhanced event rate, 4-flavour mixing can lead to
substantial decrease in the number of events, depending on the mixing
parameters.

Furthermore, we have performed a simple-minded analysis of the
systematic errors involved due to uncertainties in the supernova model
parameters. We have shown that the double ratio that quantifies the
relative variation due to mixing in water and heavy water detectors is
largely insensitive to variations in the supernova model parameters
(temperature and luminosity) used.

We have also briefly discussed the neutral current events in heavy water.
These signals may facilitate determination of the overall normalisation
of the supernova neutrino spectra.

\newpage

\begin{table}[htb]
\centering
\begin{tabular}{|c|l|} \hline
No. of flavours & Neutrino flux at detector, $F_f$ \\ \hline
3 & $F_e =\epsilon^2 F_e^0 + (1 - \epsilon^2) F_{x}^0$ \\
4 & $F_e =\epsilon^2 F_e^0 + (1 - 2\epsilon^2) F_{x}^0$ \\
3 & $2 F_{x} = (1+\epsilon^2) F_{x}^0 + (1-\epsilon^2) F_e^0$ \\
4 & $2 F_{x} = (4\epsilon^2)F_{x}^0 + (1-2\epsilon^2) F_e^0$ \\
\hline
\end{tabular}
\caption{$\nu_e$, $\nu_x = \nu_\mu, \nu_\tau$ fluxes at the detector
in the extreme adiabatic limit for 3- and 4-flavour mixing.}
\label{tab:adnu}
\end{table} % table 1

\begin{table}[htb]
\centering
\begin{tabular}{|c|l|} \hline
No. of flavours & {Antineutrino flux at detector, $F_f$} \\ \hline
3 & $F_{\bar{e}} = (1-\epsilon^2)c^2_\omega F_{\bar{e}}^0
+ (s^2_\omega + \epsilon^2 c^2_\omega) F_{x}^0$ \\
4 & $F_{\bar{e}} = (1-2\epsilon^2)c^2_\omega F_{\bar{e}}^0
+ (2\epsilon^2) F_{x}^0$ \\
3 & 2$F_{\bar{x}} = (1+c^2_\omega -\epsilon^2c^2_\omega) F_{x}^0
	+(s^2_\omega +\epsilon^2c^2_\omega) F_{\bar{e}}^0$ \\
4 & 2$F_{\bar{x}} = (2 -4\epsilon^2) F_{x}^0 +
+2\epsilon^2(1-s_{2\omega}) F_{\bar{e}}^0$ \\
\hline
\end{tabular}
\caption{$\overline{\nu}_e$, $\overline{\nu}_x = \overline{\nu}_\mu,
\overline{\nu}_\tau$ fluxes at the detector for 3- and 4-flavour mixing.
Here $c_\omega = \cos\omega$, $s_\omega = \sin \omega$.}
\label{tab:adnubar}
\end{table} % table 2

\begin{table}[ht]
\centering
\begin{tabular}{|c|l|} \hline
No. of flavours & Neutrino flux at detector, $F_f$ \\ \hline
3 & $F_{e} =(1-\epsilon^2)[ (1-P_L)s_{\omega}^2 + P_L c_\omega^2] F_{e}^0$ \\
 & $+ [1- (1-\epsilon^2)((1-P_L)s^2_{\omega} + P_Lc_{\omega}^2)]F_{x}^0$ \\
4 & $F_{e} =(1-2\epsilon^2)[ (1-P_L)s_{\omega}^2 + P_L c_\omega^2] F_{e}^0 +
2\epsilon^2 F_{x}^0$ \\
3 & $2F_{x} =[1+(1-\epsilon^2)( (1-P_L)s_{\omega}^2 + P_L c_\omega^2)] 
F_{x}^0$ \\
  & $ + [1- (1-\epsilon^2)((1-P_L)s^2_{\omega} + P_Lc_{\omega}^2)]F_{e}^0$ \\
4 & $2F_{x} =2(1-2\epsilon^2)F_{x}^0 +
2\epsilon^2 \left[1+(1-2P_L)s_{2\omega}\right] F_{e}^0$ \\
\hline
\end{tabular}
\caption{Neutrino fluxes at the detector when non-adiabatic 
effects are introduced. While the transition is assumed to be 
fully non-adiabatic at the upper resonances, it is controlled by the 
jump probability $P_L$ at the lower resonance. Here $c_\omega$,
$s_\omega$ and $s_{2\omega}$ refer to $\cos\omega$, $\sin\omega$, and
$\sin 2\omega$ respectively.}
\label{tab:nonad} 
\end{table} % table 3

\begin{table}[htb]
\begin{tabular}{|c|r|rr|rr|r|rr|rr|}
 & \multicolumn{5}{c|}{Heavy water}
 & \multicolumn{5}{c|}{Water} \\   \hline
& No & \multicolumn{2}{c|}{3-flavours, $s^2_{2\omega}=$} & 
     \multicolumn{2}{c|}{4-flavours, $s^2_{2\omega}=$} & No &
     \multicolumn{2}{c|}{3-flavours, $s^2_{2\omega}=$} & 
     \multicolumn{2}{c|}{4-flavours, $s^2_{2\omega}=$} \\
 & Mixing
 & 0.96 & 0.006 & 0.96 & 0.006 & Mixing
 & 0.96 & 0.006 & 0.96 & 0.006 \\ \hline

$\nu_e \, d(p)$ &
 72 & 183 & 183 & 181 & 181 &   0 &   0 &   0 &   0 &   0 \\

$\overline{\nu}_e \, d(p)$ &
 71 &  85 &  71 &  43 &  71 & 290 & 329 & 291 & 177 & 291 \\

$\nu\, e $ &
  8 &   9 &   9 &   8 &   8 &   9 &  10 &  10 &   9 &   9 \\

$\nu\, O $ &
  5 &  29 &  26 &  24 &  26 &   5 &  32 &  28 &  27 &  29 \\

Total &
156 & 306 & 289 & 256 & 286 & 304 & 371 & 329 & 213 & 329 \\
\end{tabular}
\caption{Total number of events in 1 kton $D_2O$ with electrons
(positrons) in the final state with energy, $E_e > 5$ MeV. Listed are
the contributions from CC events on deuterons due to both $\nu_e$ and
$\overline{\nu}_e$, events due to the elastic scattering of all
flavours and antiflavours of neutrinos on electrons (labelled $\nu\,e$),
and the CC events from $\nu_e$ and $\overline{\nu}_e$ scattering on
oxygen nuclei (labelled $\nu\,O$). The results due to no-mixing, and
mixing with 3- and 4-flavours in the adiabatic case with $\epsilon =
0.08$ are shown in the three columns.  The results with 3- and
4-flavour mixing are shown for two values of $\omega$: $\omega$ large
($\sin^2 2\omega = 0.96$) and $\omega$ small ($\sin^2 2\omega =
0.006$). For comparison, the total number of events in water are also
listed for the same set of model mixing parameters.  The deuteron
target is replaced by free protons here. }
\label{tab:CCad}
\end{table}

\begin{table}[htb]
\begin{tabular}{|c|r|rr|rr|r|rr|rr|}
 & \multicolumn{5}{c|}{Heavy water}
 & \multicolumn{5}{c|}{Water} \\ \hline
& No & \multicolumn{2}{c|}{3 flavours, $s^2_{2\omega}=$} & 
     \multicolumn{2}{c|}{4 flavours, $s^2_{2\omega}=$} & No &
     \multicolumn{2}{c|}{3 flavours, $s^2_{2\omega}=$} & 
     \multicolumn{2}{c|}{4 flavours, $s^2_{2\omega}=$} \\
 & Mixing
 & 0.96 & 0.006 & 0.96 & 0.006 & Mixing
 & 0.96 & 0.006 & 0.96 & 0.006 \\ \hline

$\nu_e\, d(p)$ &
 72 & 139 &  88 &  29 &  52 &   0 &   0 &   0 &   0 &   0 \\

$\overline{\nu}_e\, d(p)$ &
 71 &  85 &  71 &  43 &  71 & 290 & 329 & 290 & 174 & 290 \\

$\nu\, e$ &
  8 &   9 &   8 &   5 &   7 &   9 &  10 &   9 &   6 &   8 \\

$\nu\, O$ &
  5 &  21 &   8 &   3 &   4 &   5 &  23 &   8 &   3 &   5 \\

Total &
156 & 254 & 175 &  80 & 134 & 304 & 362 & 307 & 183 & 303 \\
\end{tabular}
\caption{The same as Table~\ref{tab:CCad} but for a small value of
$\epsilon = 10^{-4}$ so that the upper resonance(s) is non-adiabatic.}
\label{tab:CCnonad}
\end{table}

\begin{table}[htb]
\begin{tabular}{|c|l|l|}
                        & \multicolumn{2}{l|}{Models allowed  by the
			corresponding value of $R$ measured in} \\
 & $D_2O$ & $H_2O$ \\ \hline 
$R > 1$ & $(3\omega_L)_{A,N}$, $(4\omega_L)_{A}$, $(3,4\omega_S)_{A}$
                          & $(3\omega_L)_{A,N}$ \\
$R < 1$ & $(4\omega_L)_{N}$ & $(4\omega_L)_{A,N}$ \\
$R \sim 1$ & No mix, $(3,4\omega_S)_{N}$   & No mix, $(3,4\omega_S)_{A,N}$ \\
\end{tabular}
\caption{List of neutrino models which can be discriminated by the
$D_2O$ and $H_2O$ detectors from different values of $R$. Here $R$ is
defined as the ratio of the observed number of events with electrons of
energy $E_e > 5$ MeV in the final state, to the calculated number
without mixing. The various models are specified by the number of
flavours 3 or 4, by the value of $\omega$ ($\omega_L$ and $\omega_S$
refer to $\sin^2 2\omega = 0.96, 0.006$ respectively), and by the
suffix $A$ and $N$ referring to adiabatic and non-adiabatic propagation
at the upper resonance(s), corresponding to $\epsilon$ much larger or
much smaller than $10^{-2}$ respectively.}
\label{tab:r}
\end{table}

\begin{table}[htb]
\begin{tabular}{|c|c|}
 & Models which are allowed \\ \hline
$R_D > 1$; $R_H > 1$         & $(3\omega_L)_{A,N}$ \\
$R_D > 1$; $R_H < 1$         & $(4\omega_L)_{A}$ \\
$R_D > 1$; $R_H \sim 1$      & $(3,4\omega_S)_{A}$ \\
$R_D < 1$; $R_H > 1$         & None \\
$R_D < 1$; $R_H < 1$         & $(4\omega_L)_{N}$ \\
$R_D < 1$; $R_H \sim 1$      & None \\
$R_D \sim 1$; $R_H > 1$      & None \\
$R_D \sim 1$; $R_H < 1$      & None \\
$R_D \sim 1$; $R_H \sim 1$   & No mixing, $(3,4\omega_S)_{N}$ \\
\end{tabular}
\caption{Combined predictions from supernovae signals in water and
heavy water and corresponding models with 3 and 4 flavour mixing that
are consistent with them. By ``None'' we mean none of the models of
mixing that we have considered here. The notation is the same as in
the earlier table with the ratios $R_D$ and $R_H$ referring to heavy
water and water respectively. }
\label{tab:rdh}
\end{table}

\begin{table}[htb]
\begin{tabular}{|c|c|c|c|c|}
 & \multicolumn{3}{c|}{Number of events on deuteron} &  $R_{4/0}$ \\
 $\sin^22\omega$ & No mixing & 3-flavours & 4-flavours & \\ \hline
   0.960  & 374  &   374     &   274     &    0.73 \\
   0.760  & 374  &   374     &   281     &    0.75 \\
   0.006  & 374  &   374     &   293     &    0.78 \\
\end{tabular}
\caption{NC events on deuteron with $E_\nu > 3$ MeV for different
values of $\omega$ in the adiabatic case when $\epsilon = 0.08$. While
the 3-flavour case is identical to the no-mixing case, as expected, the
4-flavour case shows a depletion in events due to loss into the sterile
channel. The ratio of the 4-flavour to the no-mixing (or 3-flavour)
case is shown in the last column.}
\label{tab:NCad}
\end{table}

\begin{table}[htb]
\begin{tabular}{|c|c|c|c|c|}
 & \multicolumn{3}{c|}{Number of events on deuteron} & $R_{4/0}$ \\
 $\sin^22\omega$ & No mixing & 3-flavours & 4-flavours & \\ \hline
   0.960  & 374  & 374       &  336       &   0.90 \\
   0.760  & 374  & 374       &  338       &   0.90 \\
   0.006  & 374  & 374       &  365       &   0.98 \\
\end{tabular}
\caption{The same as Table~\ref{tab:NCad}, with $\epsilon \sim 0$ so
that the upper resonance is non-adiabatic.}
\label{tab:NCnonad}
\end{table}

\newpage
~ % forces the space to be left on a new page.

\newpage
~

\newpage
~

\newpage
~

\newpage
~

\newpage
~

\newpage
~

\end{document}